# A critical analysis of cognitive load measurement methods for evaluating the usability of different types of interfaces: guidelines and framework for Human-Computer Interaction

Ali Darejeh[a], Nadine Marcus[a], Gelareh Mohammadi[a], John Sweller[b]

a School of computer science, Faculty of engineering, UNSW, Sydney, Australia

b School of education, Faculty of arts, design and architecture, UNSW, Sydney, Australia

Corresponding author email: ali.darejeh@unsw.edu.au

**Abstract**

Usability testing is an essential part of product design, particularly for user interfaces. To enhance the reliability of usability evaluations, employing cognitive load measurement methods can be highly effective in assessing the mental effort required to complete tasks during user testing. This review aims to provide an overview of the most suitable cognitive load measurement methods for evaluating various types of user interfaces, serving as a valuable resource for guiding usability assessments. To bridge the existing gap in the literature, a systematic review was conducted, analyzing 76 articles with experimental study designs that met the eligibility criteria. The review encompasses different methods of measuring cognitive load applicable to assessing the usability of diverse user interfaces, including computer software, information systems, video games, web and mobile applications, robotics, and virtual reality applications. The results highlight the most widely utilized cognitive load measurement methods in software usability, their respective usage percentages, and their application in evaluating the usability of each user interface type. Additionally, the advantages and disadvantages of each method are discussed. Furthermore, the review proposes a framework to assist usability testers in selecting an appropriate cognitive load measurement method for conducting accurate usability evaluations.

1. **Introduction and theoretical background**

Cognitive load theory (CLT) was introduced and developed by educational researchers to facilitate learning (Sweller, 1988). It provides instructional guidance based on human cognitive architecture and our knowledge of evolutionary psychology (Sweller, Ayres, & Kalyuga, 2011a; Sweller, Van Merriënboer, & Paas, 2019). Cognitive load (CL) is defined as the cognitive resources required to acquire a concept or learn a procedure (Sweller, 2011; Sweller, 1988). CLT emphasises the limitations of working memory when processing new information during learning or problem solving (Paas Renkl, & Sweller, 2004; Sweller, 1988; Sweller, Ayres, & Kalyuga, 2011a).

However, education is not the only domain in which CLT can be used, as mental overload can directly affect people's performance on tasks such as working with machines and devices (Engström, Johansson & Östlund, 2005; Huttunen et al., 2011). Interface usability is one of the areas that can significantly benefit from CLT, since the structure of the interface can affect software usability and user's cognitive load (Hu, Ma, & Chau, 1999; Saadé & Otrakji, 2007). Interface usability can be improved through measuring user's cognitive load and optimising the interface based on issues that affect user's cognitive load (Müller et al., 2008).

Usability is one of the central concepts of human computer interaction (HCI) that is defined as the extent to which a user can work with a product effectively, efficiently, and with satisfaction (Bevan, Carter, & Harker, 2015). Usability testing is important to evaluate if the product can be used by target users easily (Nielsen, 1994). Since in the context of software interfaces, usability is determined by considering target users' characteristics, and specific tasks required by the interface, in order to design highly usable software applications an in-depth understanding of target users and their tasks is necessary (Sharp, Preece & Rogers, 2019). Therefore, cognitive load measurement methods can be used as an efficient technique to conduct an accurate usability evaluation.

Although there are few studies that reviewed different cognitive load measurement methods in the area of computer science, to the best of our knowledge none of them listed and compared all the CL measurement methods in the context of software usability and indicated which method is more suitable for evaluating the usability of each type of user interfaces. For example, Duran, Zavgorodniaia and Sorva (2022) conducted a systematic review to identify how CLT has been used across a number of computing education research forums. Frazier, Pitts and McComb (2022) examined how cognitive load is evaluated in studies with focus on human-automated knowledge work. They compared the studies that evaluated the impacts of automation on cognitive load of operators. Kosch et al. (2023) conduct a systematic review on studies that used cognitive load measurement methods in the area of human computer interaction. Although they reviewed some usability studies that used CL measurement methods, the focus of their paper was not on identifying all the possible cognitive load measurement methods that can be used in the area of usability and identifying which one is more appropriate for each type of user interface.

In this paper, we have analysed different methods of measuring cognitive load to determine which methods are best suited to evaluate the usability of different types of user interfaces and discuss advantages and disadvantages of each method. This overview is likely to be relevant to practitioners who are interested in cognitive load measures where technology includes a user interface. Examples include artificial intelligence interfaces and applications including robots and driverless cars; E-learning software; video games and immersive technologies including applications that are designed for virtual or augmented reality; healthcare devices and manufacturing equipment; smartphone applications and websites; or flight simulators.

In the upcoming sections, we delve into the human cognitive architecture and explore the different types of cognitive load. We subsequently provide a detailed description and analysis of various measurement methods for cognitive load, encompassing subjective and objective approaches. These methods are specifically pertinent to the area of usability evaluation for diverse software interfaces. Following that, the discussion section presents a comparative analysis of these measurement methods, evaluating their strengths and limitations. This analysis is followed by the introduction of our proposed framework, which aims to assist usability testers in selecting an appropriate cognitive load measurement method for conducting precise usability evaluations. The framework takes into account various criteria to ensure the selection of the most suitable method.

### 1.1. Human cognitive architecture and cognitive load theory

Based on evolutionary psychology, information processed by the human cognitive system can be divided into biologically primary or secondary categories (Geary, 2008, 2012; Geary & Berch, 2016). The primary category consists of information that over countless generations we have evolved to acquire easily and unconsciously. An example is learning to listen to and speak our native language. Biologically secondary information consists of a category that we need for cultural reasons but require instruction and conscious effort to acquire. Learning to read and write provides an example as does learning to use a computer program. Cognitive load theory is principally concerned with biologically secondary information.

There is a specific cognitive architecture that allows humans to acquire, process and use biologically secondary information. Information can be acquired either by obtaining it during problem solving or by transmission from another person. Novel information from the environment must be processed by a limited capacity, limited duration working memory before being stored in an unlimited capacity and duration long-term memory for subsequent use. Once information is stored in long-term memory, it can be retrieved by working memory to govern action appropriate to the extant environment. Working memory has no known limits when dealing with familiar information stored in long-term memory. This cognitive architecture provides a base for cognitive load theory.

### 1.2. Different types of cognitive load

There are two types of cognitive load: intrinsic and extraneous that interact to produce the total cognitive load (Sweller, 2010).

#### 1.2.1. Intrinsic load

Intrinsic cognitive load is defined as the natural complexity level of a specific instructional topic or an entity such as a mathematical concept, software, or a device. It is fixed and cannot be changed, except by altering the entity design or knowledge level of the learner. For example, statistical analysis and 3D modelling applications are difficult in nature and it can be said that they have a high intrinsic load. Intrinsic load is related to the number of required activities to achieve a learning goal and determined by the level of element interactivity of learning materials (Sweller, 1994).

In cognitive load theory, element interactivity is an index to measure the complexity of a learning topic and depends on the learners' prior knowledge, nature of the materials and the relationship between the concepts that users should connect and process simultaneously (Chen, Kalyuga, & Sweller, 2015; Marcus, Cooper & Sweller, 1996; Sweller, 2010; Sweller, 1994). Elements with low interactivity can be learnt in isolation, without or with minimal reference to other learning elements, for example, learning the features like bold, italic and underline in MS Excel. Low interactivity elements impose a low working memory load because each element can be processed independently of every other element. However, elements with high interactivity consist of different integrated elements that cannot be learnt in isolation, for example, typing functions in MS Excel or drawing a chart (Sweller, 2010). High levels of element interactivity impose a heavy working memory load because all the elements must be processed simultaneously. Levels of intrinsic cognitive load should be optimized to ensure learning is maximized without overloading working memory.

#### 1.2.2. Extraneous load

Extraneous cognitive load is related to the difficulty level of instructional materials or the complexity level of an interface or a device as a result of the way they have been designed. Some instructional procedures increase element interactivity and so unnecessarily increase working memory load. Common examples of extraneous load include redundant information such as repeated links, menus, or decorative graphics on websites, which are non-essential to the task at hand and may in fact be distracting and use up limited cognitive resources (Bus, Takacs & Kegel, 2015; Lehmann, Hamm, & Seufert, 2019; Jin, 2012). Split attention is another common type of extraneous load, where users have to split their attention between different screens, windows or parts of an interface in order to perform a task or understand a website or interface. The user then needs to use cognitive resources to mentally integrate information that is physically separated and that are non-essential to the task at hand (Al-Shehri & Gitsaki, 2010; Schroeder & Cenkci, 2018; Sweller, Kalyuga & Ayres, 2011).

In the context of interface usability, extraneous cognitive load may refer to the design of the user interface, tooltips and software help system that are designed to guide users while working with the interface. In comparison with intrinsic load that depends on the difficulty level of the the software content extraneous load depends on the way that the interface and the instructional materials are designed and presented (Van Merriënboer & Sweller, 2005). For example, websites are generally not difficult in nature, but a poor interface design can make a website difficult to learn, or a good help system that is designed for complex software can make learning the software easier, even if the software is difficult in nature.

Designing an interface based on principles of good design and providing an appropriate help system can enhance learning performance by increasing available working memory resources and reducing extraneous load (Chandler & Sweller, 1991). When the intrinsic load is high, total cognitive load can be decreased by decreasing extraneous load. Learning is enhanced by minimizing mental resources that are allocated to deal with a user interface and teaching materials so that working-memory can use its maximum resources to deal with the instructional topic (Sweller, 2010). Therefore, when the nature of the software or the associated interface are cognitively demanding, a good user interface design or an appropriate and comprehensive help system become important ways to decrease extraneous cognitive load.

## 2. Review Methodology

In order to conduct a systematic review of literature in this domain, the researchers followed a procedure defined by Kitchenham and Charters (2007), which is one of the most complete and suitable methods for reviewing studies in computer science. We carried out this review in three main phases: a) planning of systematic mapping, b) conducting the review, and c) reporting the review. The phases of this systematic review and the related activities are shown in Figure 1.

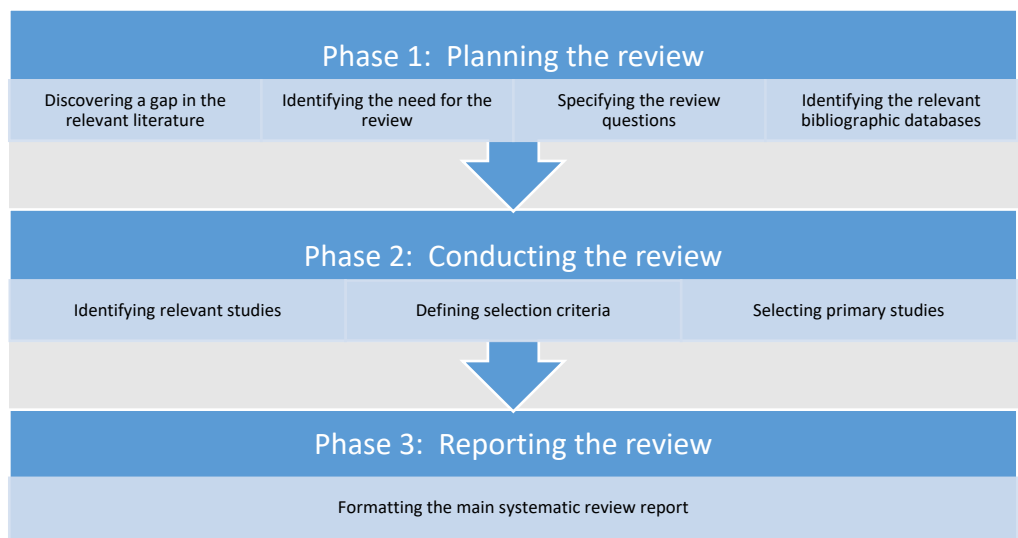

Figure 1: Phases of conducting this systematic review

### 2.1. Phase 1: Planning the review

#### 2.1.1. Discovering a gap in the relevant literature
In this step, a comprehensive search was performed on web databases to locate studies relevant to cognitive load measurement methods that are used in the area of evaluating software usability; the bibliographic databases accessed were IEEE, ACM, Science Direct, Springer Link, Wiley, Taylor & Francis, Google scholar and Scopus. To the best of our knowledge no study has conducted a comprehensive review of the existing cognitive load measurement methods in the area of software usability.

#### 2.1.2. Identifying the need for a review
Evaluating the usability issues of computer software, websites and mobile apps by considering users cognitive load is an efficient method as it considers users' cognitive structure, and it can involve both qualitative and quantitative measures. However, we are unaware of any review that summarises and compares all the possible cognitive load measurement methods that are available and can be used in the area of software usability.

#### 2.1.3. Specifying the review questions
The questions we aimed to examine for this review attempted to find and summarise all the cognitive load measurement methods that can be used in the area of software usability. The questions we aimed to investigate are listed below:
- Q1 what are the existing cognitive load measurement methods that have been effectively used in the area of software usability?
- Q2 What are the most common cognitive load measurement methods that can be used in the area of software usability?
- Q3 What types of software have been evaluated by using cognitive load measurement methods?
- Q4 What types of software have been evaluated by using each cognitive load measurement method?
- Q5 What cognitive load measurement methods have been used for evaluating the usability of each type of software?

#### 2.1.4. Identifying the relevant bibliographic databases

In order to answer our questions and find the relevant studies, we tried to select bibliographic databases that cover the majority of journals and conference papers associated with the field of computer science and gamification. The following relevant bibliographic databases were identified: ACM, IEEE, SpringerLink, ScienceDirect, Proquest, Scopus, Wiley Inter Science, and Google Scholar. The search included articles published from 2001 up to January 2023.

**2.2. Phase 2: Conducting the review**

This is the main phase of our systematic review with the purpose of selecting related studies as well as synthesising data.

**2.2.1. Identifying relevant studies**

In order to locate meaningful papers in bibliographic databases, we conducted a complete search with a combination of five categories of terms (see Table 1). The construction of search terms was based on the steps described by Brereton et al. (2007). Accordingly, Boolean OR was used for alternative spellings, synonyms or alternative terms, and Boolean AND was applied to connect the main terms.

Table 1: Search keywords

| Term 1 | | | | And | Term 2 | And | Term 3 | And | Term 4 | And | Term 5 |
|---|---|---|---|---|---|---|---|---|---|---|---|
| **NASA-TLX** | Task completion | Pupil dilation | Dual task paradigm method | | Cognitive load | | Measurement | | Usability | | Software |
| **Or** | Or | Or | Or | | Or | | Or | | Or | | Or |
| **Electroencephalography (EEG)** | Time duration | Blink rate | Performance measures | | CL | | Method | | Interface usability | | Mobile app |
| **Or** | Or | Or | Or | | | | | | | | Or |
| **Functional near-infrared spectroscopy (fNIRS)** | Mouse dynamics | Fixation | Facial expressions | | | | | | | | Website |
| **Or** | Or | Or | Or | | | | | | | | Or |
| **Magnetoencephalography (MEG)** | Linguistic Features | Saccades | Electrodermal activity (EDA) | | | | | | | | System |
| **Or** | Or | Or | Or | | | | | | | | Or |
| **Functional magnetic resonance imaging (fMRI)** | Eye Movement | Blood pressure | Heart rate Variability (HRV) | | | | | | | | Virtual reality |
| **Or** | Or | Or | | | | | | | | | |

Two additional search strategies were applied to retrieve the maximum number of relevant papers. The first strategy was reviewing the reference list of selected papers to find more related papers. The second strategy was googling the authors of selected studies to find potential related research studies. A summary of the search in bibliographic databases is presented in Table 2.

Table 2: Summary of the search in bibliographic databases

| | |
|---|---|
| **Academic databases searched** | Scopus |
| | Science Direct |
| | Wiley Inter Science |
| | ACM Digital Library |
| | Springer Database |
| | IEEE |
| | Google scholar |
| **Target items** | Journal papers |
| | Workshop papers |
| | Conference papers |
| | Book chapters |
| **Search applied to** | Title |
| | Abstract |
| | Keywords |
| | Content |
| **Language** | Papers written in English |
| **Publication period** | January 2001 up to January 2023 |

Preprint version!Preprint version!

### 2.2.2. Defining selection criteria

In order to select the primary papers, we defined the following criteria based on the purpose of our study.

**Inclusion criteria:**
1. Studies that measured users cognitive load to evaluate the usability of any type of software including computer software, websites, mobile applications, virtual reality applications and video games.
2. Studies that were published between January 2001 and January 2023.

**Exclusion criteria:**
1. Studies that measured cognitive load in another context rather than software.
2. Non-experimental papers.
3. Papers available only in the form of an abstract or presentation.
4. Papers not written in English.

### 2.2.3. Selecting primary studies

The titles and abstracts of searched papers were reviewed based on the inclusion and exclusion criteria. Every paper that met at least one of the criteria and without any of the exclusion criteria was included in the review. For papers that could not be excluded based on reading the titles and abstracts, the full texts of papers were reviewed. Through this process, 76 articles were selected.

### 2.3. Phase 3: Reporting the review

The reviewed studies were categorized into subjective versus objective, and direct versus indirect methods. At the end of each section, a table is presented that summarises studies that used each measurement method. The studies that used more than one measurement method are presented once based on the main focus of the reviewed paper under the related section. For example, if the focus of a paper is evaluating the efficiency of dual tasks paradigm, but they used other measurement methods as well, this study is only presented under the dual tasks paradigm heading.

## 3. Measuring cognitive load in usability testing

The existing methods of measuring cognitive load that can be used in evaluating software usability can be categorized into the two major dimensions: a) objectivity of measures, and b) causal relations (Brunken, Plass, & Leutner, 2003). The objectivity dimension refers to whether the measurement method uses subjective approaches such as self-reported data or objective approaches such as behavioral observation, physiological or performance measurement. The causal relations dimension categorizes methods based on whether the measured factor has a direct or indirect relation to changes in cognitive load. For example, there is a direct link between the difficulty of a user interface and cognitive load, since this difficulty is a direct effect of the cognitive load of the interface. However, an indirect link exists between errors in working with an interface and cognitive load, since incomplete knowledge can cause errors that can affect cognitive load (Brunken, Plass, & Leutner, 2003).

### 3.1. Subjective measures

Subjective methods are based on self-report approaches to collect data directly from users after working with an interface. Users can rate different factors of learning, mental effort or working experience on a five, seven, or nine-point Likert-scale questionnaire from 1 (extremely easy) to 9 (extremely difficult). In subjective techniques, it is assumed that users can reliably monitor and report their cognitive processes when working with an interface (Van Mierlo et al., 2012). These methods are widely used to measure cognitive load as they are easy to conduct, they do not interfere with the primary task and they have proven to be reliable methods to indicate the amount of cognitive load (Ayres, 2006). In order to increase the reliability of self-report methods, using multi-question assessment has been suggested (Gerjets et al., 2009).

#### 3.1.1. Subjective direct methods

The main direct method of measuring cognitive load is rating the difficulty of interface layout and functionality, after users work with an interface (DeLeeuw & Mayer, 2008; Gerjets et al., 2009; Kalyuga, Chandler, & Sweller, 1999). This scale is very sensitive in identifying differences in the cognitive load of users, however, users' knowledge, attention or task difficulty can affect the results as well (Brunken, Plass, & Leutner, 2003). Some examples of the Likert-scale questions are: 'I found the difficulty of the interface layout: Easy / Neutral / Difficult', 'I found the difficulty of working with feature x: Easy / Neutral / Difficult' and 'I found the navigation through the interface pages: Easy / Neutral / Difficult'. See Table 2 for relevant studies that used usability questionnaires.

Table 2: Studies that used usability questionnaires to evaluate usability

| Studies | Measurement method | Evaluated software | Purpose of the evaluation |
|---|---|---|---|
| **Wu, Liu & Huang (2022)** | Usability questionnaire | A block-based visual programming language development tool for learning programming | Evaluating the relationship between usability and cognitive load during learning |
| **Koć-Januchta et al. (2022)** | Usability questionnaire, cognitive load questionnaire | An e-learning system for learning biology | Evaluating the relationship between usability and |

|  | cognitive load during learning |
|---|---|

In the studies that used usability questionnaires alone or in conjunction with cognitive load questionnaires, usability questionnaires have been recognised as an efficient and reliable method of determining cognitive load to reveal usability issues.

### 3.1.2. Subjective indirect methods

In the direct subjective method, the difficulty of the interface layout and functionality are reported, however, in the indirect methods, users report their amount of mental effort devoted in understanding the interface (DeLeeuw & Mayer, 2008; Paas, Renkl, & Sweller, 2003). This method seems reliable (Brunken, Plass, & Leutner, 2003). Most researchers believe that low effort is a result of low-cognitive load and when cognitive load is high, users should expand their effort to work with the interface (Reed, Burton, & Kelly, 1985). Supporting this assumption, there is a significant correlation between difficulty rating and effort rating results (DeLeeuw & Mayer, 2008).

The choice between subjective indirect and direct cognitive load measurement methods depends on the specific research question, and the nature of the cognitive task. Subjective direct cognitive load measurement methods can provide more direct insight into the cognitive load experienced by the individual and may be easier to administer. However, the accuracy of subjective direct methods may be limited by individual differences in perception and self-report bias (Brunken, Plass & Leutner, 2003).

**NASA-TLX**

A widely used Likert scale questionnaire, that is used in measuring the mental effort of users after using a software application or a website is the NASA Task Load Index (TLX) test. This test can measure the cognitive load in various human-machine environments that engage different technologies and equipment such as control rooms, and laboratory environments (Albers, 2011). Similar to other subjective methods of measuring cognitive load, NASA-TLX cannot measure cognitive load while users are working with a website or software. The usage of the NASA-TLX occurs after completing each action and relies on users' opinion of their previous workload. For example, it can be used to evaluate the cognitive load of users after working with different tools of a software application or different parts of a website. However, since users must rely on their memory and remember the process, some details can be forgotten. Also, the default questions of NASA-TLX should be adapted based on the functionality of the software (Pachunka, 2018; Pachunka et al., 2019).

The NASA-TLX test, assesses cognitive load based on six scales consisting of: mental demand, temporal demand, physical demand, performance, effort, and frustration. It asks users the following questions:
1. how much mental effort was required? (Mental Demand);
2. how much time pressure did you feel? (Temporal Demand);
3. how much physical effort was required? (Physical Demand);
4. how hard did you work to finalize the task? (Effort);
5. how successful were you in completing the task? (Performance);
6. how disappointed, bored, or annoyed were you while completing the task? (Frustration Level) (Hart & Staveland, 1988; Schmutz et al., 2009; Pandian & Suleri, 2020).

There are many instances of the NASA-TLX test being used to evaluate usability and cognitive load. See Table 3 for relevant studies that used NASA-TLX test alone or in conjunction with the other measurement methods such as usability questionnaires and interviews.

Table 3: Studies that used the NASA-TLX test to evaluate usability

| **Studies** | **Measurement method** | **Evaluated software** | **Purpose of the evaluation** |
|---|---|---|---|
| **Clarke et al. (2020)** | NASA-TLX; Interview | Personal health record system | Usability during data entry |
| **Pachunka et al. (2019)** | NASA-TLX; Interview | Personal health record system | Usability during data entry |
| **Pollack &Pratt (2020)** | NASA-TLX | Electronic health records | Evaluating physicians' cognitive load during clinical prioritization |
| **Brendle et al. (2020)** | NASA-TLX; Task performance duration | Surgical navigation system | Evaluating if the system can facilitate spinal fusion surgery |
| **Richardson et al. (2019)** | NASA-TLX; Usability questionnaire | Electronic clinical decision support tool | Overall usability of the system for physicians |
| **Rockman et al. (2020)** | NASA-TLX ; Usability questionnaire | Cyber operations battlefield application | Performance of a system for teaching cyberspace and electromagnetic activities to military personnel. |
| **Galais, Delmas & Alonso (2019)** | NASA-TLX | Gestural interaction vs gamepad in virtual reality | Comparing cognitive load of users |

In the studies that used NASA-TLX alone or in conjunction with other cognitive load measurement methods, NASA-TLX has been recognised as an efficient and reliable method of determining cognitive load to reveal usability issues.

### 3.2. Objective measures

Although subjective measures of cognitive load are widely used by many researchers, objective methods have their own benefits. Objective methods can be categorized into brain activity measures, dual-task-paradigms, performance outcome analysis, behavioral patterns, and physiological measures other than brain activity measures. A major benefit of objective methods such as behavioral patterns and physiological measures is that they can provide a continuous measure of cognitive load that enables researchers to collect and analyze fluctuations in a stream of data over time, in contrast with the subjective self-reported techniques that only provide a few data points at the end of the usability test. They also do not rely on participants' subjective assessment of their own load.

#### 3.2.1. Direct methods

In this category, use of brain activity measures and dual task paradigm methods are common measurements.

##### 3.2.1.1. Brain activity measures

Brain activity measurement methods use neuroimaging techniques to evaluate cognitive load based on continuous brain signal measurements while users are performing a task. Neuroimaging techniques can measure brain activities when performing cognitive tasks by visualizing brain region activation (Anderson, et al., 2011; Just et al., 2001). These measurement methods have several benefits including measuring the load continuously with high sensitivity and helping researchers to distinguish stress from mental workload as they can have the same effect on learning performance. However, these measurement methods are quite intrusive for users, require a time-consuming setup, their data analysis is complex (Baldwin & Cisler, 2017) and they are not suitable for routine applications. Different techniques can be used for measuring brain activity and cognitive load including electroencephalography (EEG), functional near infrared spectroscopy (fNIRS), functional magnetic resonance imaging (fMRI), and magnetoencephalography (MEG). Since EEG and fNIRS in comparison with fMRI and MEG have simpler installation processes with portable equipment, they are more common methods.

###### 3.2.1.1.1. Electroencephalography (EEG)

One of the most popular brain activity measurement methods is EEG that is designed to capture continuous brain activity including alpha, beta, and theta waves. EEG data changes based on cognitive stimuli and working memory load (Anderson, et al., 2011). For example, when task difficulty is increased, alpha and theta bands show more activity (Gevins & Smith, 2003). EEG is performed by placing electrodes on the scalp that can measure voltage fluctuations that are related to ionic current within the brain neurons (Schomer & Da Silva, 2012). These sensors are connected via flexible fibre optic cables to a portable base, which allows researchers to measure cognitive load while participants perform a task in either a standing or walking position.
There are many instances of EEG devices being used to evaluate usability and cognitive load. See Table 4 for relevant studies that used EEG alone or in conjunction with the other measurement methods.

Table 4: Studies that used EEG devices to evaluate usability

| Studies | Measurement method | Evaluated software | Purpose of the evaluation |
|---|---|---|---|
| **Sengupta et al. (2017)** | EEG | Gaze-based virtual keyboards | Finding the efficiency of EEG in measuring usability |
| **Baig & Kavakli (2018)** | EEG; Usability Questionnaire | Different input methods in AutoCAD | Comparing users' 3D object drawing performance using speech/gesture vs. keyboard/mouse. |
| **Plazak et al. (2019)** | EEG; NASA-TLX | Image-guided surgery application | Measuring users' cognitive load |
| **Cano et al. (2021)** | EEG; Usability Questionnaire ; NASA-TLX | Interactive systems | Comparing the usability of two interactive systems |
| **Caldiroli et al. (2022)** | EEG | Online information-seeking | Measuring cognitive load |

In the studies that used EEG alone or in conjunction with other cognitive load measurement methods, EEG has been recognised as an efficient and reliable method of determining cognitive load to reveal usability issues. The findings indicated that there was no significant difference between the outcomes of the EEG and the NASA-TLX which suggests both methods can be used as reliable measures of cognitive load.

#### 3.2.1.1.2. Functional near-infrared spectroscopy (fNIRS)

fNIRS is a portable brain monitoring technology that can record changes in the cerebral blood flow of the brain. fNIRS records brain activity with the use of near-infrared spectroscopy through optical sensors that are placed on the scalp. Similar to EEG, these sensors are connected via flexible cables to a portable base, which allows researchers to measure cognitive load while participants perform a task in either a standing or walking position (Ferrari & Quaresima, 2012). See table 5 for relevant studies that used fNIRS alone or in conjunction with other measurement methods.

Table 5: Studies that used fNIRS devices to evaluate usability

| Studies | Measurement method | Evaluated software | Purpose of the evaluation |
|---|---|---|---|
| **Lamb et al. (2018)** | fNIRS | Virtual reality's science education application | Virtual reality's effect on student's cognitive load |
| **Karim et al. (2012)** | fNIRS | Video game | Visualise the load on different parts of the brain |
| **Kitabata et al. (2017)** | fNIRS | Touchscreen devices | Evaluating interface usability |
| **Hirshfield et al. (2011)** | fNIRS | Different types of software applications and video games | Measuring usability and users' cognitive load |

Studies that used fNIRS to measure cognitive load, indicate it is as an efficient and reliable method of determining cognitive load to reveal usability issues.

#### 3.2.1.1.3. Magnetoencephalography (MEG) and functional magnetic resonance imaging (fMRI)

Several studies have indicated that both these technologies can be used effectively to measure cognitive load. Since no studies as far as we are aware have been used to evaluate interface usability and since bulky equipment would probably prevent such use, these techniques will not be discussed.

### 3.2.1.2. Dual task paradigm methods

The dual-task-paradigm is another direct, objective method of measuring cognitive load based on the limited capacity of cognitive resources that can be allocated to different aspects of solving a task (Padilla et al. 2019). The dual-task method can be used with two different approaches. The first approach is adding a secondary task to a primary task to induce a memory load. In this approach the focus is on the primary task, and it is expected that performance in the primary task will be decreased with a dual-task condition in comparison with a single-task condition (Park & Brünken, 2015). For example, the primary task can be opening different pages of a website as fast as possible, and the secondary task can be clicking on a red spot on the screen. Increasing the time duration of opening website pages can be an indicator of increasing cognitive load (CL).

The second approach uses a secondary task to determine the cognitive load imposed by the primary task. Thus, the focus is on the secondary task and based on the amount of load that the primary task imposes, performance of the secondary task will be changed. For example, the primary task can be working with a specific part of an interface and the secondary task can be clicking on a red spot on the screen. Decreasing the number of clicks can be an indicator of increasing cognitive load (Haji et al., 2015; Martin, 2014; Park & Brünken, 2015; Schoor, Bannert, & Brünken, 2012).

For usability testing with the use of the dual task paradigm method, the primary task is commonly working or learning with the interface, and the secondary task is a visual observation task such as remembering a letter or a word, asking participants to press a keyboard's key or move the mouse and click a button based on changing the colour of a small window or the appearance a letter or a shape on the screen (DeLeeuw and Mayer, 2008; Schmutz et al., 2009; Van Nuland, 2017; Lai & McMahan, 2020).

One of the most popular dual task paradigm methods that is used in measuring cognitive load of users while working with a website is called the tapping task (Albers, 2011). In this method, users are asked to tap (one tap per second) with either their hand or foot while working with a website and the evaluator counts the number of taps by recording a video, recording the tapping sound, or connecting some hardware to the hands or foot of the users to count the number of taps automatically (Olive, 2004). Tapping is a secondary task to the main task which is working with the website, and puts an extra load on users. Based on the difficulty of different parts of the website, the number of taps can be either increased or decreased which can be an indicator of increasing or decreasing cognitive load.

For example, to evaluate the cognitive load of users while working with a website, we can use the tapping method by asking users to use one of their hands to control the mouse and the other hand to tap on the desk. If users face any usability issue such as an inability to find the desired feature, it is expected that the number of taps will be decreased as they should expend more mental effort to work with the website. After finding the desired feature, the number of taps should return to the normal baseline rate (Miyake, Onishi, & Poppel, 2004).

Since the dual task paradigm methods such as the tapping method, measure cognitive load continuously, they can reveal different types of usability issues such as those that cannot be measured with the other methods of measuring cognitive load such as difficulty scale questionnaires that only measure cognitive load at the conclusion of the task. It can even help to find minor usability issues that may not significantly overload users, however, discovering them can help to optimise the design (Albers, 2011). The main drawback of dual-task methods is that the secondary task may influence participants' performance on the primary task (Van Mierlo et al., 2012), and so some researchers prefer not to use it. See Table 6 for relevant studies that used the dual task paradigm alone or in conjunction with other measurement methods.

Table 6: Studies that used Dual task paradigm to evaluate usability

| Studies | Measurement method | Evaluated software | Purpose of the evaluation |
|---|---|---|---|
| **Albers (2011)** | Dual task paradigm NASA-TLX | Two websites | Evaluating the usability |
| **King (2019)** | Dual task paradigm Task performance duration | An email application | Evaluating the usability |
| **Gwizdka (2010)** | Dual task paradigm | A Wikipedia like website | Evaluating the usability |
| **Giraud, Thérouanne & Steiner (2018)** | Dual task paradigm NASA-TLX | Different versions of a website | Effect of filtering irrelevant, redundant information on usability |
| **Shelton et al. (2021)** | Dual task paradigm NASA-TLX | An ambient display (ubiquitous computing sub-class) | Evaluating users' cognitive load |
| **Na (2021)** | Dual task paradigm NASA-TLX | Information seeking on the web | Evaluating user CL during query reformulation |
| **Lai & McMahan (2020)** | Dual task paradigm | Virtual reality applications | Usability of 3 walking metaphors: Human Joystick, Scaled Walking, Walking-In-Place |
| **Schmutz et al. (2009)** | Dual task paradigm NASA-TLX | Online shopping websites | Evaluating the usability |
| **Van Nuland (2017)** | Dual task paradigm Test performance Task performance duration | Interactive e-learning tool | Comparing user CL while using two different versions |

Although, most of the studies that are indicated in Table 6 showed that dual task paradigm can be used as an effective method of measuring cognitive load, the last two studies (Schmutz et al., 2009; Van Nuland, 2017) did not find any significant difference between the cognitive load of users while using a dual task paradigm. Therefore, based on the bulk of reported studies, it can be concluded that the dual-task paradigm can be an effective measure of usability although 2 studies failed to find an effect.

### 3.2.2. Indirect methods

In this category, performance measures, behavioural measures and physiological measures are the common measurement methods.

#### 3.2.2.1. Performance measures

One of the most common methods of measuring cognitive load is analyzing performance outcomes by calculating correct answers, error rates and the time duration of performing a task. This method is indirect because it depends on mental processing speed and retrieval which can be affected by cognitive load. In order to compare the usability of two versions of an interface with the use of this method, the same tasks should be assigned to users while they are working with the interfaces and based on the number of correct answers, mistakes, or time duration, a performance mark is calculated. Since the tasks are the same, we can expect that the differences in performance outcomes are a reflection of mental load that is induced by the interface design (Antonenko & Niederhauser, 2010; Brunken, Plass, & Leutner, 2003; DeLeeuw, & Mayer, 2008; Mayer, 2001; Paas, Ayres, & Pachman, 2008; Van Mierlo et al., 2012). For example, in order to evaluate the usability of the drawing chart feature of Microsoft Excel compared to the same feature using Apple Numbers, the task completion rate, and error rate of both drawing chart features can be measured and compared.

Time-on-task is another performance measurement factor that can be calculated based on the amount of time that users spend completing different tasks while working with different elements of an interface (Brunken, Plass, & Leutner, 2003; Cranford et al., 2014). When users spend more time on a specific part of an interface, it can be used to indicate a higher cognitive load (Antonenko & Niederhauser, 2010; Brunken, Plass, & Leutner, 2003; Khawaja, Ruis, & Chen, 2007). For example, in the context of multimedia systems, navigation speed and errors that increase the time duration of the target task completion can be considered to be a result of cognitive load (Astleitner & Leutner, 1996; Yin et al., 2008). See Table 7 for relevant studies that used the performance measurement method alone or in conjunction with other measurement methods.

Table 7: Studies that used performance measurement methods to evaluate usability

| Studies | Measurement method | Evaluated software | Purpose of the evaluation |
|---|---|---|---|
| **Huang, Eades & Hong (2009)** | Test performance mark; Task performance duration; Mental effort questionnaire | Graph visualizations | Effectiveness of different methods of visualizations |
| **Da Costa et al. (2013)** | Test performance mark; Task performance duration; questionnaire | Online survey tool for self-assessment of diet and physical activity | Evaluating the usability |
| **Rhodes et al. (2019)** | Test performance mark; Task performance duration | Different modules of a neurological self-assessment platform | Evaluating the usability |
| **Holden et al. (2020)** | Test performance mark; Task performance duration; Usability questions | Mobile app to inform elderly users about risks and benefits of medications | Evaluating the usability |

Overall, based on the study findings, lower error rates, higher task completion rates and less time duration for performing different tasks while using an interface, have been found to be indicators of a lower cognitive load and higher usability.

3.2.2.2. **Behavioral measures**

Behavioral measurement methods evaluate cognitive load based on users' behaviour while they are performing a task with a user interface. The most common behavioural measures that are used for usability purposes are evaluating mouse dynamics and linguistic features.

3.2.2.2.1. **Mouse dynamics**

Mouse dynamics are one of the behavioural measures of cognitive load that are supported based on our human motor system. Mouse dynamics rely on different mouse movement attributes including speed, direction, action, distance and time (Ahmed & Traore, 2007). Based on these attributes, higher level features such as acceleration, angular velocity, scroll wheel activity, clicks/double clicks, drag-and-drop operations, point-and-click operations, and silence can be measured (Ahmed & Traore 2007; Gamboa & Fred 2004; Hocquet et al., 2004; Pusara & Brodley, 2004; Jorgensen & Yu 2011). These features are aggregated and measured in distinct time periods for analysis (Grimes & Valacich, 2015). There are different applications such as Mouse Tracker, and Mousotron that can be used to measure mouse movement data.
Several studies showed that mouse dynamic attributes decrease with increasing cognitive load, since users have less resources to perform the tasks as load increases and so have less resources left to move the mouse (Grimes & Valacich, 2015; Khawaji, et al., 2014; Rheem, Verma, & Becker, 2018). In particular, if the mouse movements are redundant to the task at hand, by increasing cognitive load the movement will be decreased. However, if the mouse movement is used to complete the main tasks, the movements may increase when increasing the complexity level of the task as participants require more effort to find the solution and complete the task, and the mouse movements may support this goal. Whether mouse movement increase or decrease as cognitive load changes, will depend on the purpose of the movements to the task at hand. See Table 8 for relevant studies that used mouse dynamics in conjunction with other measurement methods.

Table 8: Studies that used mouse dynamics to evaluate usability

| **Studies** | **Measurement method** | **Evaluated software** | **Purpose of the evaluation** |
|---|---|---|---|
| **Kortum & Acemyan (2016)** | Mouse movement data; Test performance mark; Usability questionnaire | Software application | Evaluating the usability |
| **Darejeh, Marcus & Sweller (2021)** | Mouse movement data; Test performance mark; Task performance duration | MS Access | Evaluating users' cognitive load |
| **Darejeh, Marcus & Sweller (2022)** | Mouse movement data; Test performance mark; Task performance duration | MS Access | Evaluating users' cognitive load |
| **Grimes & Valacich (2015)** | Mouse movement data; Test performance mark; Task performance duration | Web applications | Measuring users' cognitive load |
| **Hibbeln et al. (2014)** | Mouse movement data; Task performance duration | Online insurance claim forms | Measuring users' cognitive load |
| **Khawaji et al. (2014)** | Mouse movement data; Mental effort questionnaire | Chat application | Measuring users' cognitive load |
| **Fuller et al. (2020)** | Mouse movement data; Test performance mark; | Patient safety dashboard | Comparing the usability of two versions |

|  | Task performance duration |
|---|---|

Mouse dynamic measures have been recognised as an efficient and reliable method of determining cognitive load to reveal usability issues.

#### 3.2.2.2.2. Linguistic Features

One of the indirect subjective methods of measuring cognitive load that can be used in evaluating usability is analysing the complexity level of users spoken language including sentence length, number of words, punctuations, syllables, use of pauses, repetitive words, corrections, complexity and the comprehensibility level of the words. Linguistic complexity is measured by two major factors known as syntactic complexity and semantic difficulty. Syntactic complexity observes the sentence length, which is the best indicator of language complexity. Semantic difficulty analyses the use of words, their lengths (syllables and letters) and their structure (Lennon & Burdick, 2004).

Different studies showed that there will be different linguistic patterns based on the complexity level of the task and the associated cognitive load. Some have indicated in high-load situations, participants' speech rate, amplitude, speech energy, and variability will be increased (Brenner et al., 1985; Lively et al., 1993). Others have reported peak intonation and pitch range patterns (Kettebekov, 2004; Lively et al., 1993) and pitch variability (Wood et al., 2004) are related to high cognitive load. Also, by increasing the cognitive load level, participants may use significantly longer sentences with more complex structures that can make comprehension more difficult (Khawaja, Chen & Marcus, 2014). Khawaja, Chen and Marcus (2012) also found that as the task complexity increased, so teams collaborate more and their language patterns change as reflected by an increase in speech, use of longer sentences, more disagreement, and using more plural pronouns such as "we" and "they". Since linguistic methods can measure the load in real time, it can be an efficient method to find different usability issues, especially if software users need to communicate and speak with each other. See Table 9 for relevant studies that used linguistic features alone or in conjunction with the other measurement methods.

Table 9: Studies that used linguistic features to evaluate usability

| Studies | Measurement method | Evaluated software | Purpose of the evaluation |
|---|---|---|---|
| **Khawaja, Chen & Marcus (2014)** | Linguistic features; Subjective rating questionnaire | Bushfire simulator application | Measuring users' cognitive load |
| **Khawaja, Chen & Marcus (2012)** | Linguistic features including pronouns | Bushfire simulator application | Measuring collaborative cognitive load |
| **Tuček, Mount & Abbass (2012)** | Linguistic features | Sudoku games | Measuring users' cognitive load |
| **Oviatt (2006)** | Linguistic features; Number of errors; Task performance duration | Different versions of the same interface for teaching maths | Evaluating the usability |

The results of these studies indicate that linguistic features can be used as reliable measures of cognitive load. With an increase in the difficulty level of a task, the median intensity of speech decreases, and the pitch increases as well. Also, by increasing cognitive load the ability to produce complex sentences may be decreased as working memory is busy with the other tasks. Moreover, as the cognitive load of a collaborative task increases, participants work together more as reflected by changed speech patterns.

#### 3.2.2.2.3. Eye Movements

Eye movement can be simply measured with the use of cameras and eye tracking devices without the need to attach anything to users that can make it uncomfortable for them and decrease reliability of the results. Different studies have proved the efficiency of measuring eye movements in the process of usability testing and designing adaptive E-learning systems (El Haddioui, 2019). Some researchers found that when self-report questionnaires cannot show an accurate result in the usability evaluation, eye tracking methods can be used as an alternative and reliable solution (Schiessl, 2003; Wang et al., 2020).

Eye movement measures can be categorised into behavioural (voluntary) including fixations and saccades and physiological (involuntary) including pupil dilation and blink rate that will be discussed in the physiological measures section (Chen et al., 2016, Pfleging et al., 2016, Rudmann, McConkie, & Zheng, 2003).

##### 3.2.2.2.3.1. Eye Movement Fixations

Fixation is a behavioural (voluntary) eye movement measurement which refers to a situation when the eyes remains still over a period of time, from 200 milliseconds to several seconds, focussed onto an area of interest (AOI). In this measurement, the interface element that is relevant for the current cognitive activity is identified based on the gaze direction (Rudmann, McConkie, & Zheng, 2003). Increasing the fixation duration on an interface element can show that users needed more attention as the understanding of the interface was difficult for them. It can be an indication of increased processing in working memory and high cognitive load (Chen et al., 2011). See Table 10 for relevant studies that used the fixations measurement method alone or in conjunction with other measurement methods.

Table 10: Studies that measured fixations to evaluate usability

| Studies | Measurement method | Evaluated software | Purpose of the evaluation |
|---|---|---|---|
| **Zardari, Hussain & Arain (2021)** | Eye Movements (Fixation); Usability questionnaire; Test performance mark; Task performance duration | E-learning portal | Evaluating the usability |
| **Guerberof, Moorkens & Brien (2021)** | Eye Movements (Fixation); Test performance mark; Task performance duration; Satisfaction questionnaire | Microsoft Word | Evaluating the usability |
| **Realpe-Muñoz et al. (2018)** | Eye Movements (Fixation) | E-voting systems | Evaluating users' cognitive load |
| **Wang et al. (2014)** | Eye Movements (Fixation); Task performance duration | An online shopping website | Users' visual attention and cognitive load |
| **Chynał, Szymański & Sobecki (2012)** | Eye Movements (Fixation); Task performance duration | A mobile application | Evaluating the usability |
| **Bruneau et al. (2002)** | Eye Movements (Fixation); EDA | Freeserve, AOL, TISCALI and BTopenworld websites | Evaluating the usability |
| **Cheng &Wei. (2018)** | Eye Movements (Fixation); EEG | Music player software | Optimizing the interface |
| **Huang & Zhang (2022)** | Eye Movements (Fixation); usability questionnaire | E-learning | Measuring cognitive load of elderly users |
| **Calvo et al. (2022)** | Eye Movements (Fixation); Test performance mark; Task performance duration | Climate service visualizer tool | Evaluating usability to redesign the system |

The results showed that the fixation measurement method can be used as a reliable method to measure the usability and cognitive load of users.

### 3.2.2.2.3.2. Saccades

Another behavioural (voluntary) eye movement measurement involves the use of a saccade that refers to when eyes shift between two locations (from one fixation to another) voluntarily. The most common measurement method for saccades is observing the patterns of scan paths which is defined as the number of gaze fixations at different AOIs. Increasing saccade velocity from one interface element to another one indicates a higher cognitive load because a user is searching for the desired interface element, which demonstrates that the user cannot easily interact with the interface. Also, a decrease in saccade velocity can be an indication of tiredness (Chen et al., 2011; Barrios et al., 2004). See Table 11 for relevant studies that used the saccades measurement method alone or in conjunction with other measurement methods.

Table 11: Studies that used saccades to evaluate usability

| Studies | Measurement method | Evaluated software | Purpose of the evaluation |
|---|---|---|---|
| **Renshaw et al. (2003)** | Eye Movements (Saccades+ fixation); Error rate; Task performance duration | A management information system | Evaluating users' cognitive load |
| **Keskin et al. (2020)** | Eye Movements (Saccades+ fixation); EEG; Task performance duration | Google map | Evaluating the usability |
| **Ehmke & Wilson (2007)** | Eye Movements (Saccades + fixation) | bbc.co.uk and thetrainline.com | Evaluating the usability |

In these studies, the saccades method has been recognized as an efficient and reliable method of determining cognitive load to reveal usability issues.

### 3.2.2.3. Physiological measures

Physiological measures are a group of cognitive load measurement methods that are based on brain activities and our physiological reactions. By increasing mental processing, cortical activity causes a small nervous response within the body that can affect pupil dilation, blink rate, heart rate, blood pressure, facial muscles, and electrodermal activity (Zagermann, Pfeil & Reiterer, 2016).

#### 3.2.2.3.1. Pupil Dilation and blink rate

In the category of physiological cognitive load measures, analysing pupil dilation and blink rate are two of the most reliable approaches and are used by many, as they can be simply measured with the use of cameras and eye tracking devices without the need to attach anything to users that can make it uncomfortable for them and decrease reliability of the results.

As explained in the previous section, eye movement measures can be categorised into behavioural (voluntary) including fixations and saccades and physiological (involuntary) including pupil dilation and blink rate (Chen et al., 2016, Pfleging et al., 2016, Rudmann, McConkie, & Zheng, 2003). In the next sections the physiological measures of eye movement are discussed.

##### 3.2.2.3.1.1. Pupil dilation

Pupil dilation is a physiological (involuntary) eye movement measurement that is affected by cognitive processes. With increased cognitive load, users' pupils dilate and by decreasing the difficulty level of the task or towards the end of a task, pupils diameter size decreases (Chen et al., 2011; Klingner, Kumar, & Hanrahan, 2008; Zagermann, Pfeil, & Reiterer, 2016, Porta, Ricotti, & Perez, 2012; Rafiqi et al., 2015; Rudmann, McConkie, & Zheng, 2003). However, it should be considered that in addition to cognitive processes, pupils diameter size can be changed based on the brightness of the environment. Pupils' diameter size increases when the environment becomes darker to obtain more light and decreases when the environment becomes brighter. Therefore, controlling the brightness of the environment and display brightness are critical factors that can affect pupils' size and affect the reliability of the usability test results. In order to get more accurate results when testing a user interface based on pupils' size, the calibration value for the display brightness should be subtracted from the measured pupils' size (Liu, Zheng, & Zhou, 2019). See Table 12 for relevant studies that used pupil dilation measurement methods.

Table 12: Studies that used pupil dilation measurement method to evaluate usability

| Studies | Measurement method | Evaluated software | Purpose of the evaluation |
|---|---|---|---|
| **Dobhan, Wüllerich & Röhner (2022)** | Eye Movements (Pupil dilation + fixation); NASA–TLX; Test performance mark | Mobile ERP systems | Measuring usability |
| **Majooni, Akhavan & Offenhuber (2018)** | Eye Movements (Pupil dilation + fixation) | A learning management platform | Measuring usability |
| **Jiang et al., (2018)** | Eye Movements (Pupil dilation); Task performance duration; Usability questions | Ventilators' user-interfaces | Evaluating usability |

In these studies, pupil dilation has been recognised as an efficient and reliable method of determining cognitive load to reveal usability issues.

##### 3.2.2.3.1.2. Blink rate

Another physiological (involuntary) eye movement measurement is rate and latency of blinking which can show the attention of users. A high blink latency and a low blink rate are an indication of a high mental load (Chen et al., 2011). When users need to process the usage of an interface element that is not easily comprehensible for them and they need to devote more attention to a task, their blink rate will decrease. Therefore, the blink rate is an indication of increasing cognitive load in the area of human computer interaction (Joseph & Murugesh, 2020). See Table 13 for relevant studies that used the blink rate.

Table 13: Studies that used blink rate to evaluate usability

| Studies | Measurement method | Evaluated software | Purpose of the evaluation |
|---|---|---|---|
| **Derick et al. (2020)** | Eye Movements (Blink rate) | Video games | Evaluating users' cognitive |
| **Fowler, Nesbitt & Canossa (2019)** | Eye Movements (Blink rate+fixation) | Video games | Evaluating users' cognitive |
| **Mazur et al. (2019)** | Eye Movements (Blink rate); NASA–TLX; Test performance mark; Task performance duration | Electronic health record systems | Comparing the usability of two versions |
| **Corcoran et al. (2012)** | Eye Movements (Blink rate); Facial expressions | Video games | Evaluating users' cognitive |

Blink rate has been recognised as an efficient and reliable method of determining cognitive load to reveal usability issues.

### 3.2.2.3.1.3. Increasing the reliability of the eye movement measurement methods

When measuring cognitive load using eye movement measurement methods, there are three factors including individual interaction, social interaction, and environmental factors that should be taken into account (Jetter, Reiterer, & Geyer, 2014).

**Individual interaction**

Individual interaction describes the way users interact with a system interface including menus, content, and tools. Based on the system design and the activity that users perform with the system, they will be required to focus on a specific part of the interface which can influence users' fixations. Thus, interface layout and the visual presentation of content can also influence users' fixation location, duration, and rate without changing their cognitive load. Also, the interface structure as well as the task at hand can influence saccadic eye movements including length, velocity and angle of saccades (Zheng, 2017). For example, while users perform a visual analytic task in spreadsheet applications, they need to switch between different tools, worksheets, and charts which can affect fixation and saccadic eye movements. Another example is the difference between eye movements when users are searching for information by reading a text versus checking an image.

The other factor that can influence saccadic eye movements and fixations without affecting users' cognitive load is the input and output technologies including mouse, keyboard and touch screen displays that are used to interact with the interface. Input devices might require users to fixate on the input in addition to the visual information on a display. For example, graphic designers often need to switch between desktop displays, and digital pen and tablet to draw an artwork (Keim et al, 2008). Input and output devices can also affect pupil dilation and blinking behaviour as the luminance of the devices such as mobile phones, tablets and laptops can be changed frequently based on the brightness of the environment (Pfleging et al., 2016). Thus, in terms of individual interaction, it is important to consider the software content, the task at hand, and the input and output devices as additional elements that can influence eye movements without changing cognitive load.

**Social Interaction**

Social interaction describes the social aspects that influence the way we collaborate with the use of the system. Some systems such as online games, telecommunications applications, and e-learning platforms are inherently social, as users need to work with the system in groups and have communication and perform coordination activities with the other users. Depending on the activity, the number of users that interact with each other, and social roles of users, fixations and saccadic eye movements can be affected without changing cognitive load (Zagermann, Pfeil, & Reiterer, 2016). Switching focus between different users and interactive devices is exhausting for the eyes and consequently it can increase blinking rates and the velocity of eye movement as well as fixations and saccades. Therefore, the interference of eye movements through communication with the other users should be considered in order to increase the reliability of the usability test results when they are measured using eye movements (Zagermann, Pfeil, & Reiterer, 2016).

**Environmental factors**

Environmental factors describe the characteristics of the physical environment such as the office, lab, class or conference room where users interact with the system. Since each environment has its own characteristics such as temperature, humidity, and luminance with different types of furniture, equipment and devices, users' attention and consequently eye movements can be changed during the usability test without changing cognitive load. Thus, during a usability test the fixations should be evaluated with respect to the physical surroundings. Also, the environmental characteristics can exhaust the eye and increase blinking. For example, the environmental luminance has the highest impact on pupil size. Therefore, environmental factors should be taken into account when conducting a usability test using an eye movement measurement method as it can interfere with eye tracking measurements in several different ways (Chen et al., 2016).

### 3.2.2.3.2. Electrodermal activity (EDA)

Electrodermal activity (EDA) is one of the characteristics of the human body that causes continuous changes in the electrical properties of the skin (Boucsein, 2012; Critchley, 2002). EDA theory indicates that skin resistance varies based on the state of sweat glands in the skin and the sympathetic nervous system is responsible in controlling sweating (Martini & Bartholomew, 2001). By arousing the sympathetic nervous system, the activity of sweat glands is increased, enhancing skin conductance. Therefore, skin conductance can be measured as an indication of physiological arousal (Carlson, 2013). In other words, based on environmental factors and a person's cognitive state, emotional arousal changes can increase sweat gland activity and consequently lead to an increase in skin conductance. It should be noted that the EDA signal is representative of the intensity of an emotion but not emotion type.

EDA is measured by attaching very thin sensors around the fingers or wrist with the sensors sending the signals wirelessly to a computer, then the computer quantifies moments of significant increased arousal (iMotions, 2021). Due to the simplicity of the EDA measurement method, the EDA accuracy in detecting changes in human mental states, and the straight forward methods for quantifying EDA data, it is thus a suitable technique for measuring cognitive load during usability evaluation (Georges et al., 2017; Veilleux et al., 2020; Nourbakhsh et al., 2017).

There are two other commonly used terminologies for EDA, namely galvanic skin response (GSR) and skin conductance response (SCR) that fall under the umbrella term of EDA, and they all essentially refer to the same concept. In fact, EDA has been known as GSR and SCR historically. In the recent studies, researchers mostly used the acronym EDA, however, in older studies either GSR or SCR were used (Boucsein, 2012). See Table 14 for relevant studies that used EDA.

Table 14: Studies that used EDA devices to evaluate usability

| Studies | Measurement method | Evaluated software | Purpose of the evaluation |
|---|---|---|---|
| **Shi et al. (2007)** | EDA | Traffic control application | Evaluating users' cognitive load |
| **Barathi et al. (2020)** | EDA; Interview; Test performance mark | Video game | Evaluating users' cognitive load |
| **Nakasone, Prendinger & Ishizuka (2005)** | EDA | Video game | Evaluating users' engagement and emotion while playing the video game |
| **Mandryk, Atkins & Inkpen (2006)** | EDA; Difficulty questionnaire | Video game | Evaluating users' engagement and emotion while playing the video game |

The EDA method has been recognised as an efficient and reliable method of determining cognitive load to reveal usability issues.

### 3.2.2.3.3. Heart rate Variability (HRV) and Blood pressure

Heart rate Variability (HRV) is a measure of variation in the time interval between heartbeats, and it is an indication of changes in blood pressure and mental stress that are controlled by the autonomic nervous system (ANS) and hypothalamus in the brain (Buccelletti et al., 2009; Solhjoo et al., 2019). By increasing cognitive load, both sympathetic and parasympathetic components of the ANS increase, that impact the HRV (Goldstein et al., 2011; Solhjoo et al., 2019). In fact, the relationship between HRV and cognitive load is indirect, because by increasing cognitive load, heart rate and blood pressure increase, which can lead to a decrease in HRV (Ayres et al., 2021; Hjortskov et al., 2004). It should be noted that HRV is a more accurate and sensitive measure of cognitive load compared to heart rate or blood pressure alone, since HRV reflects the central pathway in the cardiovascular mechanism, however, the blood pressure is more influenced by the working muscle conditions (Hjortskov et al., 2004).

HRV is more successful with short-duration basic tasks such as binary decision tasks and to measure large differences in cognitive load such as differences between mentally active and mentally inactive periods (Paas, van Merriënboer & Adam, 1994). Based on several studies, using HRV in longer-lasting learning tasks can decrease the validity and sensitivity of the results (Ayres et al., 2021). Therefore, HRV is more appropriate to evaluate the location of a specific component of the interface that only has one step rather than evaluating a long task with the use of the interface.

HRV can be measured with the use of electrocardiogram (ECG), blood pressure, ballistocardiogram, and photoplethysmography (Bruser et al., 2011; Brüser, Winter, & Leonhardt, 2012). In order to evaluate cognitive load, ECG is the most common approach of measuring HRV carried out by placing electrodes on the chest skin (Forte, Favieri & Casagrande, 2019).

Most of the studies that evaluated the efficiency of the HRV method in measuring cognitive load in the context of usability testing, used HRV with a combination of other approaches such as EDA and EEG. See Table 15 for relevant studies that used HRV alone or in conjunction with other measurement methods.

Table 15: Studies that used HRV devices to evaluate usability

| Studies | Measurement method | Evaluated software | Purpose of the evaluation |
|---|---|---|---|
| **Chan et al. (2020)** | HRV, EDA, & interview | Conversational memory coach mobile application | Evaluating users' cognitive |
| **Solovey et al. (2014)** | HRV & EDA | In-vehicle user interfaces | Evaluating users' cognitive |
| **Collins et al. (2019)** | HRV & EDA | E-learning in virtual reality | Evaluating users' cognitive |
| **Rajavenkatanarayanan et al. (2020)** | HRV & EDA | Robot interfaces | Evaluating users' cognitive during robot co-operation |
| **Zhou et al. (2020)** | HRV, EEG & EDA | Surgery robotic systems | Evaluating users' cognitive |
| **Gupta et al. (2019)** | HRV, EEG & EDA | Virtual agent within a virtual reality interface | Interaction between cognitive load and users trust |

HRV in conjunction with other cognitive load measurement methods have been recognised as an efficient and reliable method of determining cognitive load to reveal usability issues. The reason HRV has not been used alone as a measure of cognitive load is that HRV alone may not provide a complete picture of cognitive load. While changes in HRV have been linked to cognitive load, they can also be influenced by other factors, such as physical activity, stress, or emotions. Moreover, different types of cognitive tasks may elicit different patterns of HRV, making it difficult to interpret HRV data without additional measures. Therefore, HRV is often used in conjunction with other cognitive load measures to provide a more comprehensive assessment of cognitive load (Bong, Fraser & Oriot, 2016).

### 3.2.2.3.4. Facial expressions

Measuring facial expression could identify the mental load based on micro-movements of facial muscles, such as frowning, which are an indication of the emotional state of an individual (Cacioppo et al., 1986). Both positive and negative emotions lead to different changes in the face muscles such as changing the form of mouth, eyebrows, and cheeks. In order to measure facial expression, the facial movements are recorded with a regular high-quality camera and the start, duration, and end of each facial movement are taken into account (Freitas-Magalhães, 2013). Commonly an angry face indicates high mental load, but neutral and happy faces indicate low mental load (Pecchinenda & Petrucci, 2016; Johanssen, Bernius & Bruegge, 2019). One of the issues in using facial expressions compared with the other measurement methods of cognitive load such as physiological and task performance is that the procedure will show less variation from high to low arousal when there is no affective interference (Hussain, Calvo & Chen, 2014). Therefore, it cannot be used to compare different variations of cognitive load during usability testing when there is not a significant difference between the mental loads.

Although, there are different studies that have used facial expression in order to evaluate the usability of different interfaces, there are not many studies in the area of evaluating facial expression in measuring cognitive load for usability test purposes. See Table 16 for relevant studies that used facial expression alone or in conjunction with other measurement methods.

Table 16: Studies that used facial expression

| Studies | Measurement method | Evaluated software | Purpose of the evaluation |
|---|---|---|---|
| **Xu et al. (2018)** | Facial expressions; Test performance mark; Task performance duration; Satisfaction questionnaire | In-vehicle information systems | Evaluating the usability |
| **Dargent et al. (2019)** | Facial expressions; Eye Movements (Pupil dilation); EDA | Web interfaces | Measuring users' cognitive load |
| **Guran, Cojocar & Dioşan (2020)** | Facial expressions | An E-learning application | Evaluating the usability for children |
| **Machado et al. (2018)** | Facial expressions; Task performance duration; Mouse movement data; Test performance mark | Lung auscultation–sound software | Evaluating the usability |
| **Thüring & Mahlke, 2007** | Facial expressions; Usability questionnaire; EDA; HRV; Test performance mark; Task performance duration | Menu and button design of a video player | Comparing the usability of two different versions |

Facial expression alone or in conjunction with other cognitive load measurement methods have been recognised as an efficient and reliable method of determining cognitive load to reveal usability issues.

## 4. Results

In this section, we will explore the answers to the review questions based on the reviewed literature.

**Q1 What are the existing cognitive load measurement methods that have been effectively used in the area of software usability?**
The review revealed that almost all the cognitive load measurement methods including NASA-TLX, Usability questionnaire, Electroencephalography (EEG), Functional near-infrared spectroscopy (fNIRS), Dual task paradigm method, Performance measures (task completion and time duration), Mouse dynamics, Linguistic Features, Fixations, Saccades, Pupil dilation, Blink rate, Electrodermal activity (EDA), Heart rate Variability (HRV), Blood pressure, and Facial expressions can and have been used to evaluate software usability. The only methods that have not been used in any of the existing studies are Magnetoencephalography (MEG) and Functional magnetic resonance imaging (fMRI).

**Q2 What are the most common cognitive load measurement methods that can be used in the area of software usability?**
Based on the current studies, performance measures (task completion and time duration) are the most used measurement methods, used within 20% of the studies. This is followed by, Electrodermal activity (EDA), NASA-TLX, Usability questionnaire, Dual task paradigm method, and Eye Movements (Fixations) are the most common usability evaluation approaches. Together, they are used in more than 57% of the studies. Facial expressions, Blink rate, Saccades, Pupil dilation, Linguistic Features, and Functional near-infrared spectroscopy (fNIRS) are the least common usability evaluation methods that are used in only 23% of the studies. Figure 2 shows the percentage of each evaluation method used in the reviewed studies.

Preprint version!

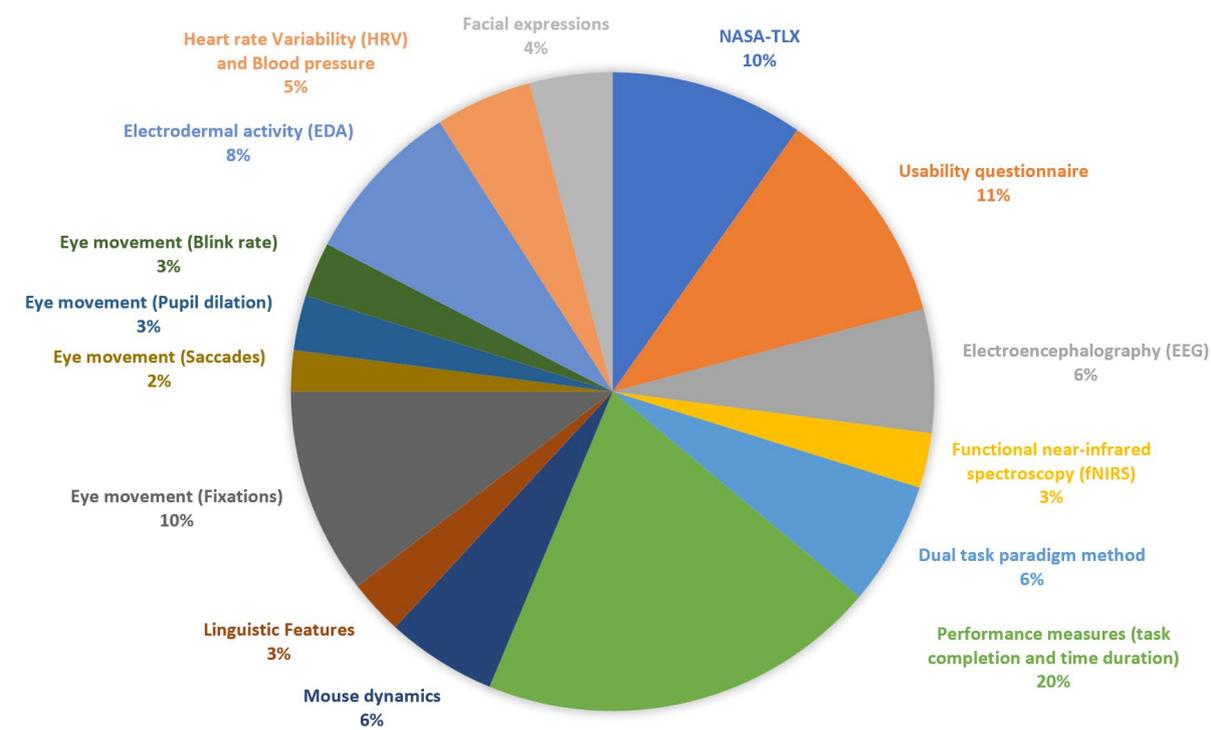

Figure 2: the percentage of each cognitive load measurement method that is used to evaluate usability

**Q3 What types of software are evaluated by using cognitive load measurement methods?**
Different types of software have been evaluated by including cognitive load measurement techniques. As it can be seen from Figure 3, evaluating the usability of websites using cognitive load measurement methods are the most common software groups with 26%, in the next rank are the video games and productivity software with 11% each, in the third rank are the information systems, E-Learnings and virtual reality applications with 8% each, and in the last rank are mobile apps, in-vehicle user interfaces and simulators with 7%, 4% and 4% respectively.

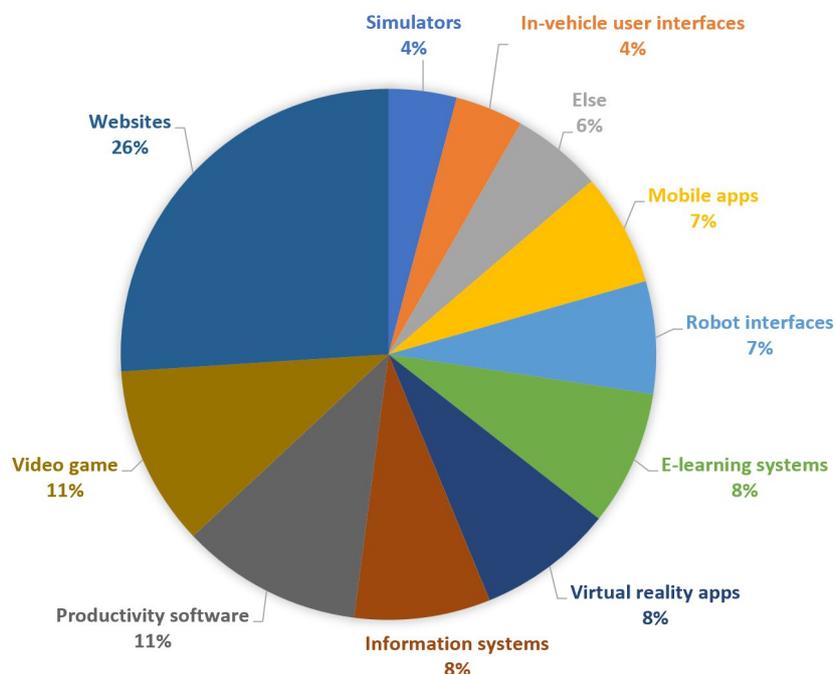

Figure 3: The percentage of each type of user interfaces that are evaluated using cognitive load measurement methods

**Q4 What types of software have been evaluated by using each cognitive load measurement method?**
In order to answer this research question, different types of software that have been evaluated with the use of each cognitive load measurement method are grouped together. Some of the main things we found include that usability questionnaires are most commonly used to evaluate the usability of productivity software and E-learning systems; NASA-TLX is most commonly used to evaluate the usability of websites, and information systems; EEG is most commonly used to evaluate the usability of productivity software; the dual task paradigm is mostly used to evaluate the usability of websites; Performance measures are mostly used to evaluate the usability of websites, mobile apps, productivity software, information systems and e-learnings;

Mouse dynamics are mostly used to evaluate the usability of websites and productivity software; Linguistic Features are mostly used to evaluate the usability of simulators; Fixations are most commonly used to evaluate the usability of websites, productivity software, e-learnings, and mobile apps; Blink rate is most commonly used to evaluate the usability of video games; EDA is most commonly used to evaluate the usability of video games and robot interfaces; HRV are most commonly used to evaluate the usability of robot interfaces; and Facial expressions are most commonly used to evaluate the usability of productivity software. For more details see Figure 4.

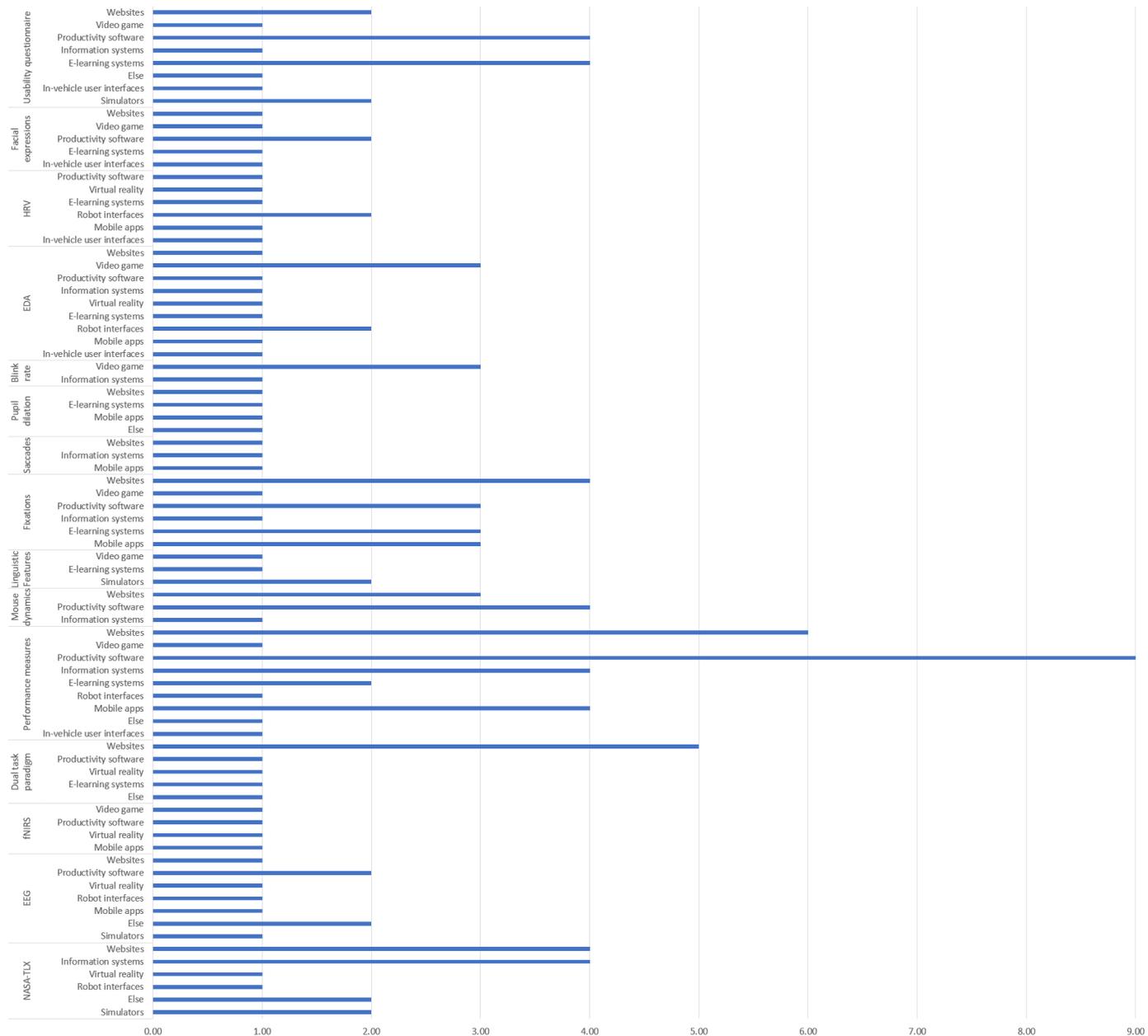

Figure 4: Different types of software that are evaluated using each cognitive load measurement method

**Q5 What cognitive load measurement methods have been used for evaluating the usability of each type of software?**
In order to answer this research question, different types of cognitive load measurement methods are grouped together for each software category. As some highlights we found that Websites are evaluated more often using NASA-TLX, Dual task paradigm method, Performance measures, Mouse dynamics, and Fixations; Video games are mostly evaluated using Blink rate, and EDA; Productivity software are evaluated using mostly Usability questionnaires, Performance measures, and Mouse dynamics; Information systems are evaluated more often using NASA-TLX and Performance measures; E-learning systems are evaluated more often using Usability questionnaire, and Fixations; Mobile apps are mostly evaluated using Performance measures and Fixations. For more details see Figure 5.

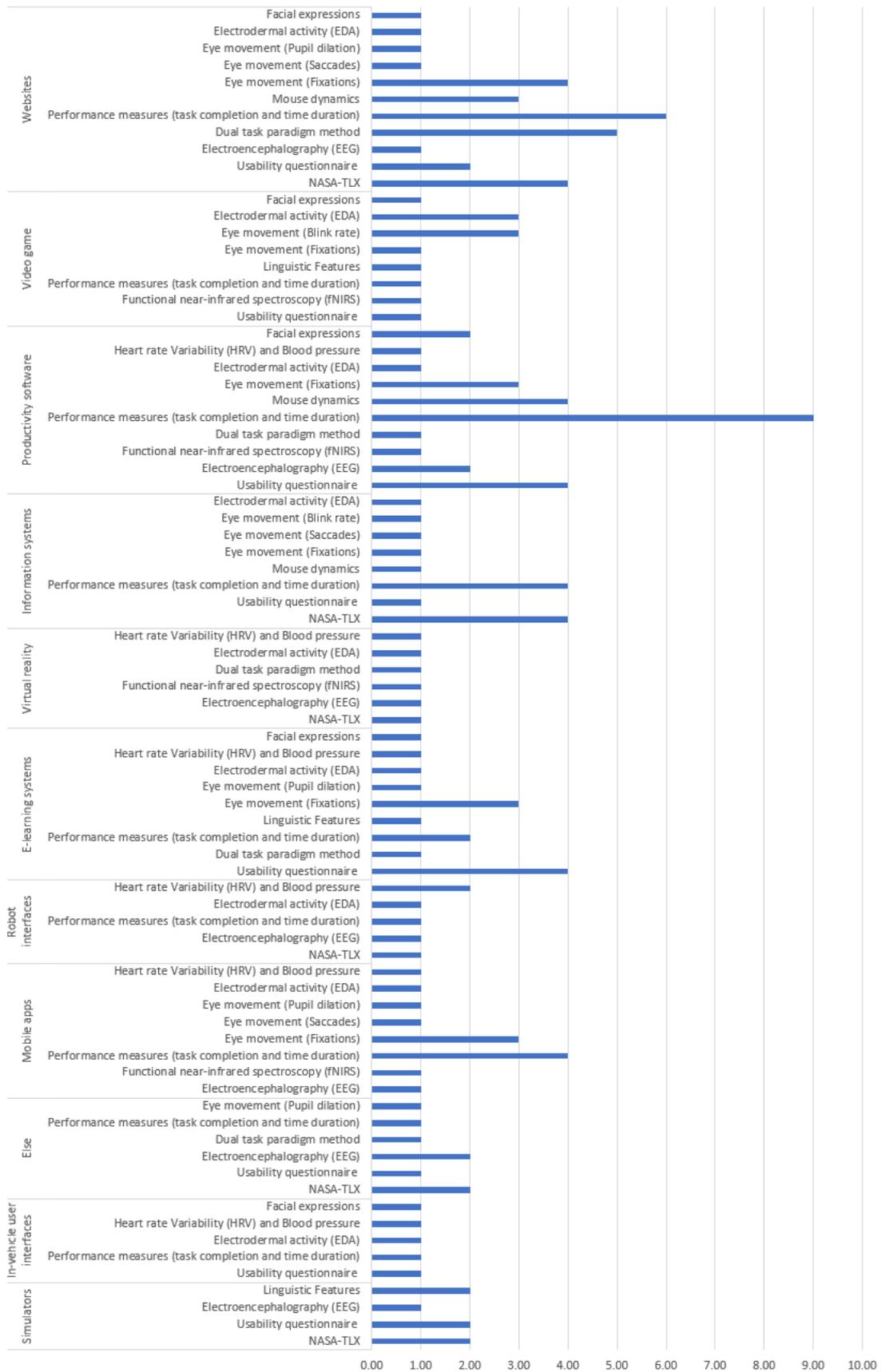

Figure 5: cognitive load measurement methods that have been used for evaluating the usability of each type of software

## 5. Discussion

In previous sections, different cognitive load measurement methods that can be used in usability evaluations were presented and in this section advantages and disadvantages of each method will be discussed. A summary of the measurement methods can be seen in Table 18.

Table 18: Classification of cognitive load measurement's methods

|  | Direct measurement | Indirect measurement |
| --- | --- | --- |
| **Subjective methods** | • Rating the difficulty of the teaching materials. | • Rating the amount of mental effort devoted to understanding the learning materials. |
| **Objective methods** | • Brain activity measures using EEG and fNIRS.<br>• Dual task paradigm such as the tapping method. | • Analyzing performance outcomes using time, error rates, correct answers.<br>• Analyzing behavioral patterns including mouse dynamics, and linguistic features.<br>• Measuring physiological factors using eye movements, pupil dilation, electrodermal activity, heart rate variability and facial expressions. |

*Subjective cognitive load measurement methods* evaluate cognitive load by asking users to complete a Likert-Scale questionnaire at the end of the test. These methods are widely used in usability evaluations as they can be easily conducted with minimum cost and the results can be easily interpreted (Ayres, 2006; Paas, 1992; Paas et al., 2016). However, since subjective methods are based on self-reported questionnaires, such as NASA TLX, that collect data directly from users at the end of the usability test, the reliability of these methods theoretically may be lower compared to the objective methods. For example, if during the usability test, users feel that their computer knowledge will be judged and it is not prestigious if they consider the interface as complex or express that they spent a great amount of mental effort, there is a chance that they may underestimate the load and potentially bias the results by providing inaccurate answers to the questionnaire. Also, even if users try to answer the questions accurately, since the questionnaire is used at the end of the usability test rather than during the test, users might forget some usability issues that they faced due to recency effects and may not answer the questions accurately. Therefore, these methods may be more appropriate for obtaining users overall opinion about the usability of the interface rather than asking about the interface details. Notwithstanding, because of their ease of use, these tests are frequently used and appear to be reliable.

*Objective cognitive load measurement methods* have a more complicated measurement process and analysis compared to subjective methods. They are also more time consuming and costly to setup. The main benefit of the objective methods is providing a continuous measure of cognitive load that enables researchers to collect and analyze fluctuations in a stream of data over time, in contrast with self-reported techniques that only provide a few data points at the end of the usability test, but are less subjective.

*Brain activity measurement* methods evaluate cognitive load directly based on the brain signals generated continuously while users are working with an interface. The main benefit of these measurement methods is measuring the load continuously and with high sensitivity, however, they are quite intrusive, require a time-consuming setup process and their data analysis is complex (Baldwin & Cisler, 2017). Since EEG and fNIRS in comparison with fMRI and MEG have easier installation process with portable equipment, they are more common methods to study users' cognitive load while working with an interface. Overall, the difficulty and inconvenience of working with brain activity measurement devices can limit the use of these techniques (Van Mierlo et al., 2012). Therefore, when considering the inconvenience of these devices for users, these measurement methods should not be used for a long usability testing process. Furthermore, these measurements should not be used for virtual reality-based applications, since the VR headset bands can prevent placing the required cables on all parts of the scalp which can decrease the reliability of the results.

*Dual task paradigm methods* evaluate cognitive load directly by adding a secondary task that is commonly a visual observation task such as clicking or pressing a key based on changing the colour, or a letter appearing on the screen while engaged in the primary task to measure memory load (Schmutz et al., 2009; Van Nuland, 2017). One of the popular dual task paradigm methods for the purpose of usability testing is a tapping task that is conducted by asking users to tap with their hand while working with an interface (Albers, 2011). A decrease in the number of taps can show a high cognitive load and reveal a usability issue. The main benefit of this method is that there is no need for any equipment to measure the load. However, a secondary task may influence participants' performance on the primary task or affect participants' concentration and change their attention from focusing on the primary task to the secondary task (Van Mierlo et al., 2012). This method is commonly used for long usability testing such as evaluating the usability of different parts of a website, but it has its drawbacks due to its potential impact on the primary task performance.

*Performance measures* evaluate the usability indirectly based on the completion rate of the tasks and completion speed (Antonenko & Niederhauser, 2010; Brunken, Plass, & Leutner, 2003; Cranford et al., 2014; Holmqvist et al., 2011). Higher task completion rates and higher performing speeds are indications of lower cognitive load. Since performance measurement methods can be used to evaluate a process and make a comparison between two different designs, they are more suitable to be used when we want to evaluate the usability in terms of the process of performing a task with two different interfaces.

*Behavioural measurement methods* evaluate cognitive load indirectly based on users' behaviour while they are working with an interface by evaluating either mouse dynamics or linguistic features.

*Mouse dynamics* measure cognitive load by evaluating different mouse movement attributes including speed, direction, action, distance and time (Ahmed & Traore 2007) and calculating the time intervals between mouse movements. Depending on the nature of the task, by increasing cognitive load, mouse movement can be either increased or decreased (Kortum & Acemyan, 2016; Darejeh, Marcus & Sweller, 2021; Grimes & Valacich, 2015; Khawaji, et al., 2014; Rheem, Verma, & Becker, 2018). Since users interact with most of the user interfaces with the use of mouse, the main benefit of this method is that users can simply perform the usability testing tasks without the need to do any extra work or use any additional equipment. However, the reliability of this method can be decreased if users do not concentrate sufficiently or there are some distracting factors in the test environment (Chen et al., 2016). Therefore, in order to compensate for the potential issues of the environmental factors, it is suggested that this method be used in conjunction with other measurement methods as the reliability of using mouse movements may be low.

*Linguistic Features* measure cognitive load by analysing the level of users' spoken language as there are different linguistic patterns based on the complexity level of the task and the associated cognitive load. In high-load situations, participants' speech rate, amplitude, speech energy, variability, sentences length, and speech complexity are increased (Brenner et al., 1985; Lively et al., 1993). Although, this measurement method can be used easily without the need of any equipment, it cannot be used to evaluate the usability of all types of interfaces. This method can be used only for applications where users require verbal communication with each other such as testing the usability of telecommunication applications or online video games.

*Physiological measurement methods* evaluate cognitive load based on physiological responses indirectly by measuring eye movements, pupil dilation, electrodermal activity, heart rate variation, blood pressure, and facial muscles. They can be intrusive to collect and measure, and specialist equipment is often needed.

*Eye Movements* can enable us to measure cognitive load based on fixations, saccades, pupil dilation and blink (Chen et al., 2016, Pfleging et al., 2016). Increasing the fixation duration on an interface element (Chen et al., 2011), increasing saccade velocity from one interface element to another (Chen et al., 2011; Barrios et al., 2004), increasing pupils' diameter size (Zagermann, Pfeil, & Reiterer, 2016; Rafiqi et al., 2015) and increasing and decreasing blink latency rate are all indications of a high cognitive load (Chen et al., 2011). Different studies have demonstrated the accuracy of eye measurement methods in the process of usability evaluation (Cheng & Wei., 2018; El Haddioui, 2019; Jiang et al., 2018; Wang et al., 2020). Also, since most of the tracking devices can measure eye movement by special cameras without attaching any device to users, these measurement methods can be used for the entire duration of a usability test. Despite all the benefits of eye-tracking methods, studies showed that different factors such as depression, illumination, or tiredness can affect the degree of pupil dilation or eye movement, affecting the reliability of usability test results (Siegle, Steinhauer, & Thase, 2004).

*Electrodermal activity (EDA)* measures skin conductance based on the state of sweat glands (Boucsein, 2012). Increases in sweat gland activity and consequently increases in skin conductance is an indication of high cognitive load (Boucsein, 2012; Critchley, 2002; Carlson, 2013; Nourbakhsh et al., 2017). Since, EDA is usually measured by attaching some sensors around the fingers or the wrist, it can affect the working performance of users while using a mouse in case of finger sensors. Therefore, this method should not be used in conjunction with the mouse movement measurement method as it can affect the reliability of the mouse movement data. Even if the sensors are attached to the opposite hand, they can still make users uncomfortable while working with the other hand. This method is accurate if the sensors are accurate enough, and so the method is used in a wide variety of studies for measuring usability and cognitive load (Nourbakhsh et al., 2017). The main drawback of this method is that to analyse the results, we may need to develop machine learning-based algorithms to predict the cognitive load.

*Heart rate Variability (HRV)* is a measure of variation in time intervals between heartbeats and is an indication of changes in blood pressure. Increasing cognitive load increases heart rate and blood pressure which can decrease HRV (Ayres et al., 2021; Hjortskov et al., 2004). Although, HRV is a relatively accurate measurement method, it is not a comfortable method for users as it is measured using ECG by placing electrodes on the chest skin, or with some newer ECG devices, on the hand and wrist. Also, since HRV is more successful with short-duration tasks (Ayres et al., 2021), this method is more appropriate for evaluating the location of a specific component of an interface that only requires one step rather than evaluating performance in more complex tasks.

*Facial expressions* evaluate the mental load based on micro-movements of facial muscles (Cacioppo et al., 1986). Facial movements can be recorded with a regular camera and commonly an angry face indicates a high mental load, but neutral and happy faces indicate a lower mental load (Pecchinenda & Petrucci, 2016; Johanssen, Bernius & Bruegge, 2019). Although, the measurement method and the interpretation of facial movement results are very simple, this method is not very accurate and shows limited variation from high to low arousal when there is no affective interference (Hussain, Calvo & Chen, 2014). Therefore, this method should be used in conjunction with the other methods of usability evaluation especially when there is no significant difference between mental loads.

Overall, among all the measurement methods of cognitive load, self-reported questionnaires, performance measures, dual task paradigm, facial expressions, linguistic features and mouse movements measurement methods are the least obtrusive in terms of data collection and analysis. There is no need for expensive equipment or an external device that is connected to users. Also, for data analysis, there is no need for specific software in contrast with brain activity, electrodermal activity, and heart rate variability measures. Finally, we should bear in mind that the outcome of all the indirect measurement methods either subjective or objective might be the result of motivational or attentional factors rather than cognitive load (Brünken et al., 2002). So, it is essential to design methods that can distinguish between different factors effectively.

Based on the advantages and disadvantages of each measurement method, it is important to choose an appropriate measurement method depending on the context of use. In order to measure users cognitive load while working with desktop interfaces including software or websites, almost all the mentioned methods can be used, including self-reported questionnaires, performance measures, mouse movements, dual task paradigm, facial expressions, eye movements, pupil dilation, ECG, HRV, EDA, blood pressure and linguistic features. However, for mobile and tablet applications the methods including measuring mouse movements, facial expressions, eye movements, and pupil dilation are not suggested. Since facial expressions, eye movements, and pupil dilation are measured with the use of cameras that are located in front of users' eyes, it is difficult to install the cameras, as commonly users hold phones and tablets with their hands and their head is down preventing an image from being captured by a camera. However, eye-tracking devices that are in the form of glasses can be used to evaluate the usability of mobile interfaces. In the future, mobile device cameras may allow this procedure but current technology does not presently support this well, and privacy issues would need to be carefully considered. Moreover, mouse movement cannot be measured in mobile device touch screen displays, as there is no mouse cursor.

In order to measure cognitive load while users are interacting with virtual reality interfaces, robots, manufacturing equipment, health care and smart home devices, measurement methods that do not prevent users from moving around or restrict their body movement are suggested, as commonly users interact with these devices while standing, and they need to have enough freedom to move their body. Therefore, the methods that need to connect a wire to users or restrict them to only sit or stand in a specific location or look in a specific direction are not recommended. The methods including facial expressions, eye movements, pupil dilation, and ECG, are thus not appropriate for the above-mentioned contexts, with the exception of device models that can provide a wireless sensor. Based on this review we proposed a framework that can guide usability testers to choose appropriate method of cognitive load measurement methods to conduct an accurate usability evaluation. Figure 6 shows our proposed framework for choosing appropriate cognitive load measurement method in the context of usability. This framework is supported by Brunken, Plass, & Leutner (2003); Schmutz et al. (2009); and Kosch et al. (2023).

Based on this framework for evaluating usability, we can measure the cognitive load of users during or after the usability test using either objective or subjective measurement methods. If the cognitive load is measured during the usability test using objective measurement methods, we can employ methods including EEG, fNIRS, Dual task paradigm, Performance measures, mouse dynamics, linguistic features, Fixations, Saccades, Pupil dilation, Blink rate, EDA, HRV, and facial expressions. However, if the cognitive load is measured after the usability test using subjective measurement methods, we can only use the Usability questionnaire and NASA-TLX. Additionally, we can measure cognitive load using either direct subjective/objective or indirect subjective/objective cognitive load measurement methods. The only direct subjective method is the Usability questionnaire, and the only indirect subjective method is NASA-TLX. Direct objective methods include EEG, fNIRS, and Dual task paradigm, while indirect objective methods include Performance measures, mouse dynamics, linguistic features, Fixations, Saccades, Pupil dilation, Blink rate, EDA, HRV, and facial expressions.

When measuring cognitive load using direct objective measurement methods, we have two options: using a device or measuring without using any device. Without using any device, we can utilize the dual task paradigm, and with the use of a device, we can choose between EEG or fNIRS. Furthermore, we can measure cognitive load using indirect objective measurement methods either with the use of a device or without using any device. When measuring without using any device, we can employ Performance measures, Mouse dynamics, and Linguistic Features. On the other hand, when using a device, we can utilize Fixations, Saccades, Pupil dilation, Blink rate, Facial expressions, EDA, and HRV. Finally, when using a direct objective measurement method with a device, we can use a device attached to users such as EEG, or an external device like fNIRS. Similarly, when using indirect objective measurement methods with a device, we can use a device attached to users like EDA and HRV, or external devices like Fixations, Saccades, Pupil dilation, Blink rate, and Facial expressions.

Preprint version!

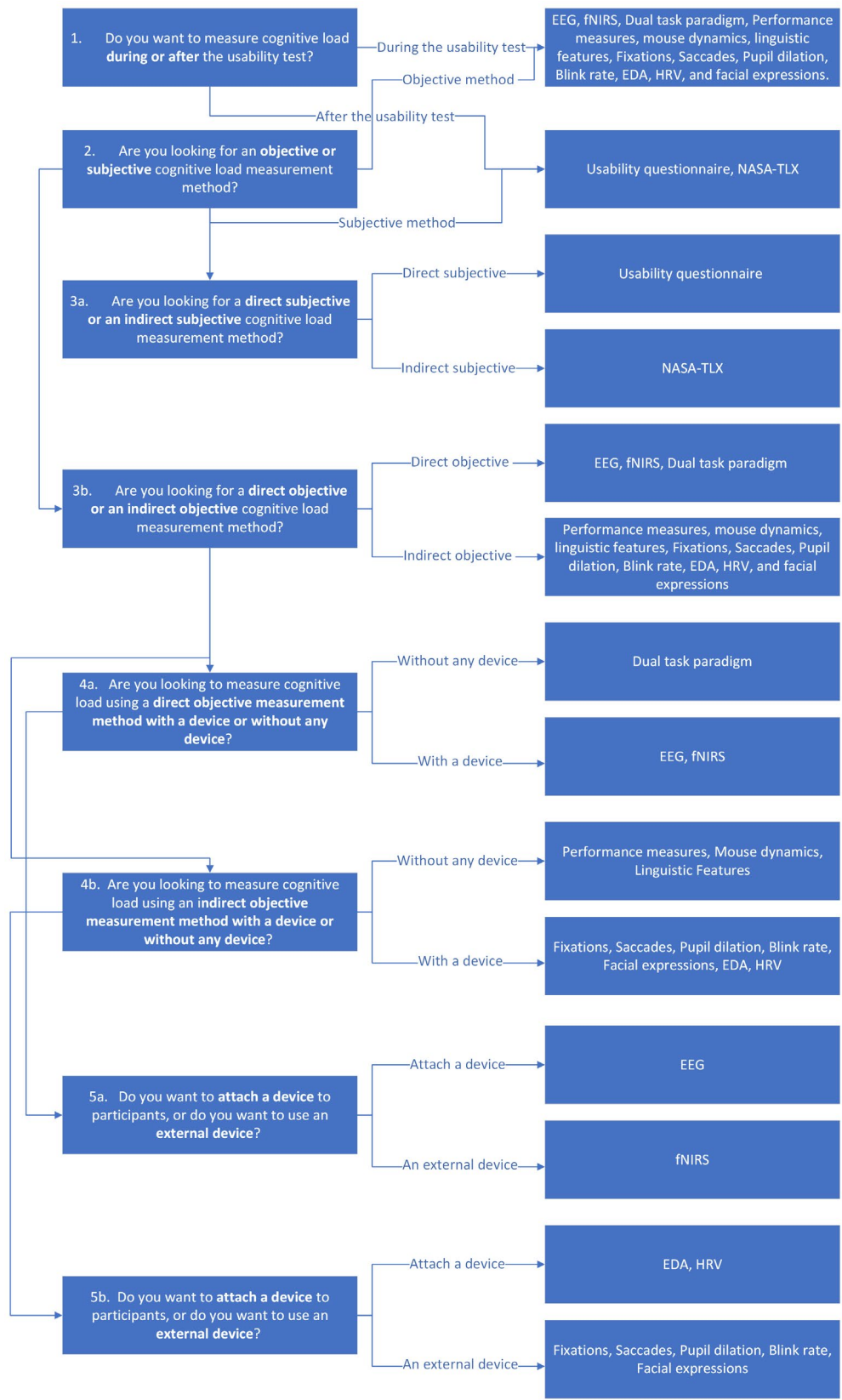

Figure 6: The framework for choosing appropriate cognitive load measurement method in the context of usability

## 6. Conclusions

In this paper a critical review and analysis of different cognitive load measurement methods that can be used to evaluate the usability of interfaces is provided. As a summary, a cognitive load measurement method that will be used during a usability test process should have two characteristics: a) it should not create any discomfort for users, and b) it should measure the cognitive load continuously, not just at the end of the session. Although, methods such as brain activity measurement procedures including EEG, and physiology measures such as heart rate variability and electrodermal activity can measure cognitive load continuously and objectively, they can cause stress or distraction especially when the experiment duration is long (Van Mierlo et al., 2012). Furthermore, the information from eye-tracking methods can be affected by personal or environmental factors such as tiredness or brightness (Siegle, Steinhauer, & Thase, 2004). Hence, proper consideration has to be made when using such methods.

By considering the advantages and disadvantages of different measurement methods, it can be concluded that the most suitable methods to measure cognitive load during a usability test are performance on task, time on task, dual task paradigms especially the tapping method, mouse dynamics, facial expressions, and linguistic features if applicable. These methods measure the mental load continuously without causing any discomfort or distraction for the participants. They also do not need expensive equipment and can be used by non-expert people as well. In addition to the above methods, since self-reported difficulty or mental effort questionnaires can provide valuable information and do not create any discomfort for the participants, they can be used as a complementary method at the completion of a usability test.

Finally, in order to conduct an accurate usability test, a combination of subjective and objective cognitive load measurement methods should be utilised to compensate for the shortcomings of each method and increase the reliability of the results.

**References**

Ahmed, A. A. E., & Traore, I. (2007). A new biometric technology based on mouse dynamics. IEEE Transactions on dependable and secure computing, 4(3), 165-179.

Albers, M. J. (2011, October). Tapping as a measure of cognitive load and website usability. In Proceedings of the 29th ACM international conference on Design of communication (pp. 25-32).

Al-Shehri, S., & Gitsaki, C. (2010). Online reading: A preliminary study of the impact of integrated and split-attention formats on L2 students' cognitive load. ReCALL, 22(3), 356-375.

Anderson, E. W., Potter, K. C., Matzen, L. E., Shepherd, J. F., Preston, G. A., & Silva, C. T. (2011). A User Study of Visualization Effectiveness Using EEG and Cognitive Load. Computer Graphics Forum, 30(3), 791–800. https://doi.org/10.1111/j.1467-8659.2011.01928.x

Antonenko, P. D., & Niederhauser, D. S. (2010). The influence of leads on cognitive load and learning in a hypertext environment. Computers in Human Behavior, 26(2), 140-150.

Astleitner, H., & Leutner, D. (1996). Applying Standard Network Analysis to Hypermedia Systems: Implications for Learning. Journal of Educational Computing Research, 14(3), 285–303. https://doi.org/10.2190/W2GB-05NTVJRN-PGY9

Ayres, P. (2006). Using subjective measures to detect variations of intrinsic cognitive load within problems. Learning and Instruction, 16(5), 389–400. https://doi.org/10.1016/j.learninstruc.2006.09.001

Ayres, P., Lee, J. Y., Paas, F., & van Merriënboer, J. J. (2021). The Validity of Physiological Measures to Identify Differences in Intrinsic Cognitive Load. Frontiers in Psychology, 12.

Baig, M. Z., & Kavakli, M. (2018, December). Analyzing novice and expert user's cognitive load in using a multi-modal interface system. In 2018 26th International Conference on Systems Engineering (ICSEng) (pp. 1-7). IEEE.

Baldwin, C. L., & Cisler, D. S. (2017). Neuroergonomic Methods of Assessing Learning. In Cognitive Load Measurement and Application (pp. 240-262). Routledge.

Barathi, S. C., Proulx, M., O'Neill, E., & Lutteroth, C. (2020, April). Affect recognition using psychophysiological correlates in high intensity VR exergaming. In Proceedings of the 2020 CHI Conference on Human Factors in Computing Systems (pp. 1-15).

Barrios, V. M. G., Gütl, C., Preis, A. M., Andrews, K., Pivec, M., Mödritscher, F., & Trummer, C. (2004). AdELE: A framework for adaptive e-learning through eye tracking. Proceedings of IKNOW, 609-616.

Bevan, N., Carter, J., & Harker, S. (2015, August). ISO 9241-11 revised: What have we learnt about usability since 1998?. In International Conference on Human-Computer Interaction (pp. 143-151). Springer, Cham.

Bong, C. L., Fraser, K., & Oriot, D. (2016). Cognitive load and stress in simulation. Comprehensive healthcare simulation: Pediatrics, 3-17.

Boucsein, W. (2012). Electrodermal activity. Springer Science & Business Media.

Brendle, C., Schütz, L., Esteban, J., Krieg, S. M., Eck, U., & Navab, N. (2020, October). Can a Hand-Held Navigation Device Reduce Cognitive Load? A User-Centered Approach Evaluated by 18 Surgeons. In International Conference on Medical Image Computing and Computer-Assisted Intervention (pp. 399-408). Springer, Cham.

Brenner, M., Shipp, T., Doherty, E., & Morrissey, P. (1985). Voice measures of psychological stress: Laboratory and field data. In I. Titze & R. Scherer (Eds.), Vocal fold physiology, biomechanics, acoustics, and phonatory control (pp. 239–248). Denver, CO: Denver Center for the Performing Arts.

Bruneau, D., Sasse, M. A., & McCarthy, J. (2002, April). The Eyes Never Lie: The Use of Eye Tracking Data in HCI Research.

Brunken, R., Plass, J. L., & Leutner, D. (2003). Direct measurement of cognitive load in multimedia learning. Educational psychologist, 38(1), 53-61.

Brünken, R., Steinbacher, S., Plass, J. L., & Leutner, D. (2002). Assessment of cognitive load in multimedia learning using dual-task methodology. Experimental psychology, 49(2), 109–109.

Bruser, C., Stadlthanner, K., de Waele, S., & Leonhardt, S. (2011). Adaptive beat-to-beat heart rate estimation in ballistocardiograms. IEEE Transactions on Information Technology in Biomedicine, 15(5), 778-786.

Brüser, C., Winter, S., & Leonhardt, S. (2012). Unsupervised heart rate variability estimation from ballistocardiograms. In proceedings of the 7th International Workshop on Biosignal Interpretation (Vol. 15, pp. 1-6).

Buccelletti, E., Gilardi, E. M. A. N., Scaini, E., Galiuto, L. E. O. N., Persiani, R. O. B. E., Biondi, A. L. B. E., ... & Silveri, N. G. (2009). Heart rate variability and myocardial infarction: systematic literature review and metanalysis. Eur Rev Med Pharmacol Sci, 13(4), 299-307.

Bus, A. G., Takacs, Z. K., & Kegel, C. A. (2015). Affordances and limitations of electronic storybooks for young children's emergent literacy. Developmental Review, 35, 79-97.

Cacioppo, J. T., Petty, R. E., Losch, M. E., & Kim, H. S. (1986). Electromyographic activity over facial muscle regions can differentiate the valence and intensity of affective reactions. Journal of personality and social psychology, 50(2), 260.

Caldiroli, C. L., Gasparini, F., Corchs, S., Mangiatordi, A., Garbo, R., Antonietti, A., & Mantovani, F. (2022). Comparing online cognitive load on mobile versus PC-based devices. Personal and Ubiquitous Computing, 1-11.

Calvo, L., Christel, I., Terrado, M., Cucchietti, F., & Pérez-Montoro, M. (2022). Users' cognitive load: A key aspect to successfully communicate visual climate information. Bulletin of the American Meteorological Society, 103(1), E1-E16.

Cano, S. P., Soto, J., Acosta, L., Peñeñory, V., & Moreira, F. (2021). Electroencephalography as an Alternative for Evaluating User eXperience in Interactive Systems. In WorldCIST (1) (pp. 435-444).

Carlson, N. (2013). Physiology of Behavior. New Jersey: Pearson Education, Inc. ISBN 978-0-205-23939-9.

Chan, S. W., Sapkota, S., Mathews, R., Zhang, H., & Nanayakkara, S. (2020). Prompto: Investigating Receptivity to Prompts Based on Cognitive Load from Memory Training Conversational Agent. Proceedings of the ACM on Interactive, Mobile, Wearable and Ubiquitous Technologies, 4(4), 1-23.

Chandler, P.; Sweller, J. (1991). Cognitive Load Theory and the Format of Instruction. Cognition and Instruction. 8 (4): 293–332. doi:10.1207/s1532690xci0804_2.

Chen, F., Zhou, J., Wang, Y., Yu, K., Arshad, S. Z., Khawaji, A., & Conway, D. (2016). Robust multimodal cognitive load measurement. Cham: Springer.

Chen, O., Kalyuga, S., & Sweller, J. (2015). The Worked Example Effect, the Generation Effect, and Element Interactivity. Journal of Educational Psychology, 107(3), 689–704.

Chen, S., Epps, J., Ruiz, N., & Chen, F. (2011, February). Eye activity as a measure of human mental effort in HCI. In *Proceedings of the 16th international conference on Intelligent user interfaces* (pp. 315-318).

Cheng, S., & Wei, Q. (2018, November). Design preferred aesthetic user interface with eye movement and electroencephalography data. In Proceedings of the 2018 ACM Companion International Conference on Interactive Surfaces and Spaces (pp. 39-45).

Chynał, P., Szymański, J. M., & Sobecki, J. (2012, March). Using eyetracking in a mobile applications usability testing. In Asian Conference on Intelligent Information and Database Systems (pp. 178-186). Springer, Berlin, Heidelberg.

Clarke, M. A., Schuetzler, R. M., Windle, J. R., Pachunka, E., & Fruhling, A. (2020). Usability and cognitive load in the design of a personal health record. Health Policy and Technology, 9(2), 218-224.

Collins, J., Regenbrecht, H., Langlotz, T., Can, Y. S., Ersoy, C., & Butson, R. (2019, October). Measuring cognitive load and insight: A methodology exemplified in a virtual reality learning context. In 2019 IEEE International Symposium on Mixed and Augmented Reality (ISMAR) (pp. 351-362). IEEE.

Corcoran, P. M., Nanu, F., Petrescu, S., & Bigioi, P. (2012). Real-time eye gaze tracking for gaming design and consumer electronics systems. IEEE Transactions on Consumer Electronics, 58(2), 347-355.

Cranford, K. N., Tiettmeyer, J. M., Chuprinko, B. C., Jordan, S., & Grove, N. P. (2014). Measuring load on working memory: the use of heart rate as a means of measuring chemistry students' cognitive load. Journal of Chemical Education, 91(5), 641-647.

Critchley, H. D. (2002). Electrodermal responses: what happens in the brain. The Neuroscientist, 8(2), 132-142.

Da Costa, F. F., Schmoelz, C. P., Davies, V. F., Di Pietro, P. F., Kupek, E., & de Assis, M. A. A. (2013). Assessment of diet and physical activity of brazilian schoolchildren: usability testing of a web-based questionnaire. JMIR research protocols, 2(2), e31.

Darejeh, A., Marcus, N., & Sweller, J. (2021). The effect of narrative-based E-learning systems on novice users' cognitive load while learning software applications. Educational Technology Research and Development, 1-23.

Darejeh, A., Marcus, N., & Sweller, J. (2022). Increasing learner interactions with E-learning systems can either decrease or increase cognitive load depending on the nature of the interaction, (3), 405-437.

Dargent, T., Karran, A., Léger, P. M., Coursaris, C. K., & Sénécal, S. (2019). The Influence of Task Types on User Experience after a Web Interface Update. Proceedings of the Eighteenth Annual Pre-ICIS Workshop on HCI Research in MIS, Munich, Germany

DeLeeuw, K. E., & Mayer, R. E. (2008). A comparison of three measures of cognitive load: Evidence for separable measures of intrinsic, extraneous, and germane load. Journal of Educational Psychology, 100(1), 223- 234.

Derick, L. R., Gabriel, G. S., Máximo, L. S., Olivia, F. D., Noé, C. S., & Juan, O. R. (2020, November). Study of the user's eye tracking to analyze the blinking behavior while playing a video game to identify cognitive load levels. In (Vol. 4, pp. 1-5). IEEE.

Dobhan, A., Wüllerich, T., & Röhner, D. (2022). Eye-Tracking and Usability in (Mobile) ERP Systems. In International Conference on Enterprise Information Systems (pp. 403-423). Springer, Cham.

Duran, R., Zavgorodniaia, A., & Sorva, J. (2022). Cognitive Load Theory in Computing Education Research: A Review. ACM Transactions on Computing Education (TOCE), 22(4), 1-27.

Ehmke, C., & Wilson, S. (2007). Identifying Web Usability Problems from Eye-Tracking Data. In British Computer Society.

El Haddioui, I. (2019). Eye Tracking Applications for E-Learning Purposes: An Overview and Perspectives. Cognitive Computing in Technology-Enhanced Learning, 151-174.

Engström, J., Johansson, E., & Östlund, J. (2005). Effects of visual and cognitive load in real and simulated motorway driving. Transportation research part F: traffic psychology and behaviour, 8(2), 97-120.

Ferrari, M., & Quaresima, V. (2012). A brief review on the history of human functional near-infrared spectroscopy (fNIRS) development and fields of application. Neuroimage, 63(2), 921-935.

Forte, G., Favieri, F., & Casagrande, M. (2019). Heart rate variability and cognitive function: a systematic review. Frontiers in neuroscience, 13, 710.

Fowler, A., Nesbitt, K., & Canossa, A. (2019). Identifying Cognitive Load in a Computer Game: An exploratory study of young children. In 2019 IEEE Conference on Games (CoG) (pp. 1-6). doi: 10.1109/CIG.2019.8848064.

Frazier, S., Pitts, B. J., & McComb, S. (2022). Measuring cognitive workload in automated knowledge work environments: a systematic literature review. Cognition, Technology & Work, 24(4), 557-587.

FREITAS-MAGALHÃES, A. (2013). Facial expression of emotion: from theory to application. Leya.

Fuller, T. E., Garabedian, P. M., Lemonias, D. P., Joyce, E., Schnipper, J. L., Harry, E. M., ... & Benneyan, J. C. (2020). Assessing the cognitive and work load of an inpatient safety dashboard in the context of opioid management. Applied ergonomics, 85, 103047.

Galais, T., Delmas, A., & Alonso, R. (2019, December). Natural interaction in virtual reality: impact on the cognitive load. In Proceedings of the 31st Conference on l'Interaction Homme-Machine: Adjunct (pp. 1-9).

Gamboa, H., & Fred, A. (2004, August). A behavioral biometric system based on human-computer interaction. In Biometric Technology for Human Identification (Vol. 5404, pp. 381-392). International Society for Optics and Photonics.

Geary, D. (2008). An evolutionarily informed education science. Educational Psychologist, 43, 179-195.

Geary, D. (2012). Evolutionary Educational Psychology. In K. Harris, S. Graham, & T. Urdan (Eds.), APA Educational Psychology Handbook (Vol. 1, pp. 597-621). American Psychological Association.

Geary, D., & Berch, D. (2016). Evolution and children's cognitive and academic development. In D. Geary & D. Berch (Eds.), Evolutionary perspectives on child development and education (pp. 217-249). Springer.

Georges, V., Courtemanche, F., Sénécal, S., Léger, P. M., Nacke, L., & Pourchon, R. (2017, July). The adoption of physiological measures as an evaluation tool in UX. In International Conference on HCI in Business, Government, and Organizations (pp. 90-98). Springer, Cham.

Gerjets, P. W., Hesse, F. W. H., Scheiter, K., Eysink, T. H. S., & Opfermann, M. (2009). Learning with hypermedia: The influence of representational formats and different levels of learner control on performance and learning behavior. Computers in Human Behavior, 25(2), 360–370. https://doi.org/10.1016/j.chb.2008.12.015

Gevins, A., & Smith, M. E. (2003). Neurophysiological measures of cognitive workload during human-computer interaction. Theoretical Issues in Ergonomics Science, 4(1-2), 113-131.

Giraud, S., Thérouanne, P., & Steiner, D. D. (2018). Web accessibility: Filtering redundant and irrelevant information improves website usability for blind users. International Journal of Human-Computer Studies, 111, 23-35.

Goldstein, D. S., Bentho, O., Park, M. Y., & Sharabi, Y. (2011). Low-frequency power of heart rate variability is not a measure of cardiac sympathetic tone but may be a measure of modulation of cardiac autonomic outflows by baroreflexes. Experimental physiology, 96(12), 1255-1261.

Grimes, M., & Valacich, J. (2015). Mind over mouse: The effect of cognitive load on mouse movement behavior.

Guerberof Arenas, A., Moorkens, J., & O'Brien, S. (2021). The impact of translation modality on user experience: an eye-tracking study of the Microsoft Word user interface. Machine Translation, 35(2), 205-237. https://doi.org/10.1007/s10590-021-09267-z

Gupta, K., Hajika, R., Pai, Y. S., Duenser, A., Lochner, M., & Billinghurst, M. (2019, November). In ai we trust: Investigating the relationship between biosignals, trust and cognitive load in vr. In 25th ACM Symposium on Virtual Reality Software and Technology (pp. 1-10).

Guran, A. M., Cojocar, G. S., & Dioşan, L. (2020). A Step Towards Preschoolers' Satisfaction Assessment Support by Facial Expression Emotions Identification. Procedia Computer Science, 176, 632-641.

Gwizdka, J. (2010, August). Using Stroop task to assess cognitive load. In Proceedings of the 28th Annual European Conference on Cognitive Ergonomics (pp. 219-222).

Haji, F. A., Rojas, D., Childs, R., Ribaupierre, S., & Dubrowski, A. (2015). Measuring cognitive load: performance, mental effort and simulation task complexity. Medical education, 49(8), 815-827.

Hart, S. G., & Staveland, L. E. (1988). Development of NASA-TLX (Task Load Index): Results of empirical and theoretical research. In Advances in psychology (Vol. 52, pp. 139-183). North-Holland.

Hibbeln, M., Jenkins, J. L., Schneider, C., Valacich, J. S., & Weinmann, M. (2014, January). Investigating the effect of insurance fraud on mouse usage in human-computer interactions. In 35th International Conference on Information Systems: Building a Better World Through Information Systems, ICIS 2014.

Hirshfield, L. M., Gulotta, R., Hirshfield, S., Hincks, S., Russell, M., Ward, R., ... & Jacob, R. (2011, May). This is your brain on interfaces: enhancing usability testing with functional near-infrared spectroscopy. In Proceedings of the SIGCHI Conference on Human Factors in Computing Systems (pp. 373-382).

Hjortskov, N., Rissén, D., Blangsted, A. K., Fallentin, N., Lundberg, U., & Søgaard, K. (2004). The effect of mental stress on heart rate variability and blood pressure during computer work. European journal of applied physiology, 92(1), 84-89.

Hocquet, S., Ramel, J. Y., & Cardot, H. (2004). Users authentication by a study of human computer interactions. In Proc. Eighth Ann.(Doctoral) Meeting on Health, Science and Technology.

Holden, R. J., Campbell, N. L., Abebe, E., Clark, D. O., Ferguson, D., Bodke, K., ... & Callahan, C. M. (2020). Usability and feasibility of consumer-facing technology to reduce unsafe medication use by older adults. Research in Social and Administrative Pharmacy, 16(1), 54-61.

Holmqvist, K., Nyström, M., Andersson, R., Dewhurst, R., Jarodzka, H., & Van de Weijer, J. (2011). Eye tracking: A comprehensive guide to methods and measures. Oxford: Oxford University Press.

Hu, P. J. H., Ma, P. C., & Chau, P. Y. (1999). Evaluation of user interface designs for information retrieval systems: a computer-based experiment. Decision support systems, 27(1-2), 125-143.

Huang, T., & Zhang, J. (2022, June). Study on Experience Design of Elderly Online Learning Interface Based on Cognitive Load. In Human-Computer Interaction. User Experience and Behavior: Thematic Area, HCI 2022, Held as Part of the 24th HCI International Conference, HCII 2022, Virtual Event, June 26–July 1, 2022, Proceedings, Part III (pp. 70-86). Cham: Springer International Publishing.

Huang, W., Eades, P., & Hong, S. H. (2009). Measuring effectiveness of graph visualizations: A cognitive load perspective. Information Visualization, 8(3), 139-152.

Hussain, M. S., Calvo, R. A., & Chen, F. (2014). Automatic cognitive load detection from face, physiology, task performance and fusion during affective interference. Interacting with computers, 26(3), 256-268.

Huttunen, K., Keränen, H., Väyrynen, E., Pääkkönen, R., & Leino, T. (2011). Effect of cognitive load on speech prosody in aviation: Evidence from military simulator flights. Applied ergonomics, 42(2), 348-357.

Imotions. (2021). The iMotions EDA / GSR module. https://imotions.com/biosensor/gsr-galvanic-skin-response-eda-electrodermal-activity/

Jetter, H. C., Reiterer, H., & Geyer, F. (2014). Blended Interaction: understanding natural human–computer interaction in post-WIMP interactive spaces. Personal and Ubiquitous Computing, 18(5), 1139-1158.

Jiang, M., Liu, S., Feng, Q., Gao, J., & Zhang, Q. (2018). Usability study of the user-interface of intensive care ventilators based on user test and eye-tracking signals. Medical science monitor: international medical journal of experimental and clinical research, 24, 6617.

Jin, P. (2012). Redundancy effect. In Encyclopedia of the Sciences of Learning (pp. 2787- 2788). Springer US.

Johanssen, J. O., Bernius, J. P., & Bruegge, B. (2019, May). Toward usability problem identification based on user emotions derived from facial expressions. In 2019 IEEE/ACM 4th International Workshop on Emotion Awareness in Software Engineering (SEmotion) (pp. 1-7). IEEE.

Jorgensen, Z., & Yu, T. (2011, March). On mouse dynamics as a behavioral biometric for authentication. In Proceedings of the 6th ACM Symposium on Information, Computer and Communications Security (pp. 476-482).

Joseph, S. (2013). Measuring cognitive load: a comparison of self-report and physiological methods. Unpublished Doctoral Dissertation, Arizona State University.

Just, M., Carpenter, P., Keller, T., Emery, L., Zajac, H., & Thulborn, K. (2001). Interdependence of Nonoverlapping Cortical Systems in Dual Cognitive Tasks. NeuroImage, 14(2), 417–426. https://doi.org/10.1006/nimg.2001.0826

Kalyuga, S., Chandler, P., & Sweller, J. (1999). Managing split-attention and redundancy in multimedia instruction. Applied Cognitive Psychology, 13, 351–371.

Karim, H., Schmidt, B., Dart, D., Beluk, N., & Huppert, T. (2012). Functional near-infrared spectroscopy (fNIRS) of brain function during active balancing using a video game system. Gait & posture, 35(3), 367-372.

Keim, D., Andrienko, G., Fekete, J. D., Görg, C., Kohlhammer, J., & Melançon, G. (2008). Visual analytics: Definition, process, and challenges. In Information visualization (pp. 154-175). Springer, Berlin, Heidelberg.

Keskin, M., Ooms, K., Dogru, A. O., & De Maeyer, P. (2020). Exploring the cognitive load of expert and novice map users using EEG and eye tracking. ISPRS International Journal of Geo-Information, 9(7), 429.

Kettebekov, S. (2004, October). Exploiting prosodic structuring of coverbal gesticulation. Paper presented at the ICMI'04: 6th International Conference on Multimodal Interfaces, State College, PA.

Khawaja, M. A., Chen, F., & Marcus, N. (2012). Analysis of collaborative communication for linguistic cues of cognitive load. *Human factors*, *54*(4), 518-529.

Khawaja, M. A., Chen, F., & Marcus, N. (2014). Measuring cognitive load using linguistic features: implications for usability evaluation and adaptive interaction design. International Journal of Human-Computer Interaction, 30(5), 343-368.

Khawaja, M. A., Ruiz, N., & Chen, F. (2007). Potential speech features for cognitive load measurement. In Proceedings of the 19th Australasian conference on ComputerHuman Interaction: Entertaining User Interfaces (pp. 57-60). ACM.

Khawaji, A., Chen, F., Zhou, J., & Marcus, N. (2014, December). Trust and cognitive load in the text-chat environment: the role of mouse movement. In Proceedings of the 26th Australian Computer-Human Interaction Conference on Designing Futures: The Future of Design (pp. 324-327).

King, A. (2019). Cognitive load and its impact on usage of email applications.

Kitabata, M., Inazumi, Y., Misawa, T., Horita, Y., Sugimoto, O., & Naito, S. (2017, October). Can brain activity be used to evaluate the usability of smartphone devices?. In 2017 IEEE 6th Global Conference on Consumer Electronics (GCCE) (pp. 1-2). IEEE.

Kitchenham, B. A., & Charters, S. (2007). Guidelines for performingsystematic 1302 literature reviews in software engineering. UK: EBSE.

Klingner, J., Kumar, R., & Hanrahan, P. (2008, March). Measuring the task-evoked pupillary response with a remote eye tracker. In Proceedings of the 2008 symposium on Eye tracking research & applications (pp. 69-72).

Koć-Januchta, M. M., Schönborn, K. J., Roehrig, C., Chaudhri, V. K., Tibell, L. A., & Heller, H. C. (2022). "Connecting concepts helps put main ideas together": cognitive load and usability in learning biology with an AI-enriched textbook. International Journal of Educational Technology in Higher Education, 19(1), 11.

Kortum, P., & Acemyan, C. Z. (2016, September). The relationship between user mouse-based performance and subjective usability assessments. In Proceedings of the Human factors and Ergonomics Society Annual meeting (Vol. 60, No. 1, pp. 1174-1178). Sage CA: Los Angeles, CA: SAGE Publications.

Kosch, T., Karolus, J., Zagermann, J., Reiterer, H., Schmidt, A., & Woźniak, P. W. (2023). A Survey on Measuring Cognitive Workload in Human-Computer Interaction. ACM Computing Surveys.

Lai, C., & McMahan, R. P. (2020, November). The Cognitive Load and Usability of Three Walking Metaphors for Consumer Virtual Reality. In 2020 IEEE International Symposium on Mixed and Augmented Reality (ISMAR) (pp. 627-638). IEEE.

Lamb, R., Antonenko, P., Etopio, E., & Seccia, A. (2018). Comparison of virtual reality and hands on activities in science education via functional near infrared spectroscopy. Computers & Education, 124, 14-26.

Lehmann, J. A. M., Hamm, V., & Seufert, T. (2019). The influence of background music on learners with varying extraversion: Seductive detail or beneficial effect?. Applied Cognitive Psychology, 33(1), 85-94.

Lennon, C., & Burdick, H. (2004). The lexile framework as an approach for reading measurement and success. *electronic publication on www. lexile. com*.

Liu, Y., Zheng, B., & Zhou, H. (2019). Measuring the difficulty of text translation: The combination of text-focused and translator-oriented approaches. Target. International Journal of Translation Studies, 31(1), 125-149.

Lively, E., Pisoni, D. B., Summers, W. V., & Bernacki, R. (1993). Effects of cognitive workload on speech production: Acoustic analyses and perceptual consequences. Journal of the Acoustical Society of America, 93, 2962–2973.

Machado, A., Oliveira, A., Jácome, C., Pereira, M., Moreira, J., Rodrigues, J., ... & Marques, A. (2018). Usability of Computerized Lung Auscultation–Sound Software (CLASS) for learning pulmonary auscultation. Medical & biological engineering & computing, 56(4), 623-633.

Majooni, A., Akhavan, A., & Offenhuber, D. (2018, July). An Eye-tracking study on usability and efficiency of blackboard platform. In International Conference on Applied Human Factors and Ergonomics (pp. 281-289). Springer, Cham.

Mandryk, R. L., Atkins, M. S., & Inkpen, K. M. (2006, April). A continuous and objective evaluation of emotional experience with interactive play environments. In Proceedings of the SIGCHI conference on Human Factors in computing systems (pp. 1027-1036).

Marcus, N., Cooper, M., and Sweller, J. (1996). Understand instructions. J. Educ. Psychol. 88: 49–63.

Martin, S. (2014). Measuring cognitive load and cognition: metrics for technologyenhanced learning. Educational Research and Evaluation, 20(7-8), 592

Martini, F., & Bartholomew, E. (2001). Essentials of Anatomy & Physiology. San Francisco: Benjamin Cummings. p. 263. ISBN 978-0-13-061567-1.

Mayer, R. E. (2001). Multi-media learning. Cambridge, UK: Cambridge University Press.

Mazur, L. M., Mosaly, P. R., Moore, C., & Marks, L. (2019). Association of the usability of electronic health records with cognitive workload and performance levels among physicians. JAMA network open, 2(4), e191709-e191709.

Miyake, Y., Onishi, Y., & Poppel, E. (2004). Two types of anticipation in synchronization tapping. Acta neurobiologiae experimentalis, 64(3), 415-426.

Müller, K. R., Tangermann, M., Dornhege, G., Krauledat, M., Curio, G., & Blankertz, B. (2008). Machine learning for real-time single-trial EEG-analysis: from brain–computer interfacing to mental state monitoring. Journal of neuroscience methods, 167(1), 82-90.

Na, K. (2021). The effects of cognitive load on query reformulation: mental demand, temporal demand and frustration. Aslib Journal of Information Management.

Nakasone, A., Prendinger, H., & Ishizuka, M. (2005, September). Emotion recognition from electromyography and skin conductance. In Proc. of the 5th international workshop on biosignal interpretation (pp. 219-222).

Nielsen, J. (1994). Usability engineering. Morgan Kaufmann.

Nourbakhsh, N., Chen, F., Wang, Y., & Calvo, R. A. (2017). Detecting users' cognitive load by galvanic skin response with affective interference. ACM Transactions on Interactive Intelligent Systems (TiiS), 7(3), 1-20.

Olive, T. (2004). Working memory in writing: Empirical evidence from the dual-task technique. European Psychologist, 9(1), 32-42.

Oviatt, S. (2006, October). Human-centered design meets cognitive load theory: designing interfaces that help people think. In Proceedings of the 14th ACM international conference on Multimedia (pp. 871-880).

Paas, F. G. (1992). Training strategies for attaining transfer of problem-solving skill in statistics: a cognitive-load approach. Journal of educational psychology, 84(4), 429.

Paas, F., Ayres, P., & Pachman, M. (2008). Assessment of cognitive load in multimedia learning. Recent Innovations in Educational Technology That Facilitate Student Learning. (pp.11-35). Information Age Publishing Inc., Charlotte, NC.

Paas, F., Renkl, A., & Sweller, J. (2003). Cognitive load theory and instructional design: Recent developments. Educational psychologist, 38(1), 1-4.

Paas, F., Renkl, A., & Sweller, J. (2004). Cognitive Load Theory: Instructional Implications of the Interaction between Information Structures and Cognitive Architecture. Instructional Science, 32(1), 1–8. https://doi.org/10.1023/B:TRUC.0000021806.17516.d0

Paas, F., Tuovinen, J. E., Tabbers, H., & Van Gerven, P. W. (2016). Cognitive load measurement as a means to advance cognitive load theory. In Educational psychologist (pp. 63-71). Routledge.

Paas, F., van Merriënboer, J. J. G., and Adam, J. J. (1994). Measurement of cognitive load in instructional research. Percept. Motor Skills 79, 419–430. doi: 10.2466/pms.1994.79.1.419

Pachunka, E. (2018). Natural-Setting PHR Usability Evaluation Using Eye Tracking and NASA TLX to Measure Cognitive Load of Patients (Doctoral dissertation, University of Nebraska at Omaha).

Pachunka, E., Windle, J., Schuetzler, R., & Fruhling, A. (2019, January). Natural-setting PHR usability evaluation using the NASA TLX to measure cognitive load of patients. In Proceedings of the 52nd Hawaii International Conference on System Sciences.

Padilla, L. M., Castro, S. C., Quinan, P. S., Ruginski, I. T., & Creem-Regehr, S. H. (2019). Toward objective evaluation of working memory in visualizations: A case study using pupillometry and a dual-task paradigm. *IEEE transactions on visualization and computer graphics*, *26*(1), 332-342.

Pandian, V. P. S., & Suleri, S. (2020). NASA-TLX Web App: An Online Tool to Analyse Subjective Workload. arXiv preprint arXiv:2001.09963.

Park, B., & Brünken, R. (2015). The Rhythm Method: A New Method for Measuring Cognitive Load—An Experimental Dual-Task Study. Applied Cognitive Psychology, 29(2), 232-243.

Pecchinenda, A., & Petrucci, M. (2016). Emotion unchained: Facial expression modulates gaze cueing under cognitive load. PLoS One, 11(12), e0168111.

Pfleging, B., Fekety, D. K., Schmidt, A., & Kun, A. L. (2016, May). A model relating pupil diameter to mental workload and lighting conditions. In Proceedings of the 2016 CHI conference on human factors in computing systems (pp. 5776-5788).

Plazak, J., DiGiovanni, D. A., Collins, D. L., & Kersten-Oertel, M. (2019). Cognitive load associations when utilizing auditory display within image-guided neurosurgery. International journal of computer assisted radiology and surgery, 14(8), 1431-1438.

Pollack, A. H., & Pratt, W. (2020). Association of health record visualizations with physicians' cognitive load when prioritizing hospitalized patients. JAMA network open, 3(1), e1919301-e1919301.

Porta, M., Ricotti, S., & Perez, C. J. (2012, April). Emotional e-learning through eye tracking. In Proceedings of the 2012 IEEE Global Engineering Education Conference (EDUCON) (pp. 1-6). IEEE.

Pusara, M., & Brodley, C. E. (2004, October). User re-authentication via mouse movements. In Proceedings of the 2004 ACM workshop on Visualization and data mining for computer security (pp. 1-8).

Rafiqi, S., Wangwiwattana, C., Kim, J., Fernandez, E., Nair, S., & Larson, E. C. (2015, July). PupilWare: towards pervasive cognitive load measurement using commodity devices. In Proceedings of the 8th ACM International Conference on PErvasive Technologies Related to Assistive Environments (pp. 1-8).

Rajavenkatanarayanan, A., Nambiappan, H. R., Kyrarini, M., & Makedon, F. (2020, August). Towards a Real-Time Cognitive Load Assessment System for Industrial Human-Robot Cooperation. In 2020 29th IEEE International Conference on Robot and Human Interactive Communication (RO-MAN) (pp. 698-705). IEEE.

Realpe-Muñoz, P., Collazos, C. A., Hurtado, J., Granollers, T., Muñoz-Arteaga, J., & Velasco-Medina, J. (2018). Eye tracking-based behavioral study of users using e-voting systems. , , 182-195.

Reed, W. M., Burton, J. K., & Kelly, P. P. (1985). The Effects of Writing Ability and Mode of Discourse on Cognitive Capacity Engagement. Research in the Teaching of English, 19(3), 283–297.

Renshaw, J. A., Finlay, J. E., Tyfa, D., & Ward, R. D. (2003). Designing for visual influence: An eye tracking study of the usability of graphical management information. Human-Computer Interaction, 1, 144-51.

Rheem, H., Verma, V., & Becker, D. V. (2018, September). Use of mouse-tracking method to measure cognitive load. In *Proceedings of the human factors and ergonomics society annual meeting* (Vol. 62, No. 1, pp. 1982-1986). Sage CA: Los Angeles, CA: SAGE Publications.

Rhodes, J. K., Schindler, D., Rao, S. M., Venegas, F., Bruzik, E. T., Gabel, W., ... & Rudick, R. A. (2019). Multiple sclerosis performance test: technical development and usability. Advances in therapy, 36(7), 1741-1755.

Richardson, K. M., Fouquet, S. D., Kerns, E., & McCulloh, R. J. (2019). Impact of mobile device-based clinical decision support tool on guideline adherence and mental workload. Academic pediatrics, 19(7), 828-834.

Rockman, C. M. B., O'Shea, C. K. W., Thomson, R. H., Boyce, M. W., Vey, N. L., Geddes, J. A., ... & Colon, C. E. T. (2020) Assessing Cognitive Load and Usability for CEMA Training Using COBWebS.

Rudmann, D. S., McConkie, G. W., & Zheng, X. S. (2003, November). Eyetracking in cognitive state detection for HCI. In Proceedings of the 5th international conference on Multimodal interfaces (pp. 159-163).

Saadé, R. G., & Otrakji, C. A. (2007). First impressions last a lifetime: effect of interface type on disorientation and cognitive load. Computers in human behavior, 23(1), 525-535.

Schiessl, M., Duda, S., Thölke, A., & Fischer, R. (2003). Eye tracking and its application in usability and media research. , (2003), 41-50.

Schmutz, P., Heinz, S., Métrailler, Y., & Opwis, K. (2009). Cognitive load in eCommerce applications—measurement and effects on user satisfaction. Advances in Human-Computer Interaction, 2009.

Schomer, D. L., & Da Silva, F. L. (2012). Niedermeyer's electroencephalography: basic principles, clinical applications, and related fields. Lippincott Williams & Wilkins.

Schoor, C., Bannert, M., & Brünken, R. (2012). Role of dual task design when measuring cognitive load during multimedia learning. Educational Technology Research and Development, 60(5), 753-768.

Schroeder, N., & Cenkci, L. (2018). Spatial Contiguity and Spatial Split-Attention Effects in Multimedia Learning Environments: a Meta-Analysis. Educational Psychology Review, 30(3), 679–701. https://doi.org/10.1007/s10648-018-9435-9

Sengupta, K., Sun, J., Menges, R., Kumar, C., & Staab, S. (2017, June). Analyzing the impact of cognitive load in evaluating gaze-based typing. In 2017 IEEE 30th International Symposium on Computer-Based Medical Systems (CBMS) (pp. 787-792). IEEE.

Sharp, H., Preece, J., & Rogers, Y. (2019). Interaction design: Beyond human-computer interaction, jon wiley & sons. Inc.

Shelton, B., Nesbitt, K., Thorpe, A., & Eidels, A. (2021). Gauging the utility of ambient displays by measuring cognitive load. Cognition, Technology & Work, 23(3), 459-480.

Shi, Y., Taib, R., Ruiz, N., Choi, E., & Chen, F. (2007). Multimodal human-machine interface and user cognitive load measurement. IFAC Proceedings Volumes, 40(16), 200-205.

Siegle, G. J. J., Steinhauer, S. R. R., & Thase, M. E. E. (2004). Pupillary assessment and computational modeling of the Stroop task in depression. International Journal of Psychophysiology, 52(1), 63–76. https://doi.org/10.1016/j.ijpsycho.2003.12.010

Solhjoo, S., Haigney, M. C., McBee, E., van Merrienboer, J. J., Schuwirth, L., Artino, A. R., ... & Durning, S. J. (2019). Heart rate and heart rate variability correlate with clinical reasoning performance and self-reported measures of cognitive load. Scientific reports, 9(1), 1-9.

Solovey, E. T., Zec, M., Garcia Perez, E. A., Reimer, B., & Mehler, B. (2014, April). Classifying driver workload using physiological and driving performance data: two field studies. In Proceedings of the SIGCHI Conference on Human Factors in Computing Systems (pp. 4057-4066).

Sweller, J. (1988). Cognitive load during problem solving: Effects on learning. Cognitive science, 12(2), 257-285.

Sweller, J. (1994). Cognitive load theory, learning difficulty, and instructional design. Learning and instruction, 4(4), 295-312.

Sweller, J. (2010). Element interactivity and intrinsic, extraneous, and germane cognitive load. Educational psychology review, 22(2), 123-138.

Sweller, J. (2011). Cognitive load theory. In Psychology of learning and motivation (Vol. 55, pp. 37-76). Academic Press.

Sweller, J., Ayres, P., & Kalyuga, S. (2011a). Cognitive load theory. New York: Springer.

Sweller, J., van Merriënboer, J. J., & Paas, F. (2019). Cognitive architecture and instructional design: 20 years later. Educational Psychology Review, 31, 261-292.

Thüring, M., & Mahlke, S. (2007). Usability, aesthetics and emotions in human-technology interaction. International journal of psychology, 42(4), 253–264.

Tuček, D. C., Mount, W. M., & Abbass, H. A. (2012, November). Neural and speech indicators of cognitive load for sudoku game interfaces. In International Conference on Neural Information Processing (pp. 210-217). Springer, Berlin, Heidelberg.

Van Merrienboer, J., Sweller, J. (2005). Cognitive Load Theory and Complex Learning: Recent Developments and Future Directions. Educational Psychology Review. 17 (2) 147-177.

Van Mierlo, C. M., Jarodzka, H., Kirschner, F., & Kirschner, P. A. (2012). Cognitive load theory in elearning. In Z. Yan (Ed.), Encyclopedia of Cyberbehavior (pp. 1178- 1211). Hershey, PA: IGI Global.

Van Nuland, S. E. (2017). Examination and Assessment of Commercial Anatomical E-Learning Tools: Software Usability, Dual-Task Paradigms and Learning.

Veilleux, M., Sénécal, S., Demolin, B., Bouvier, F., Di Fabio, M. L., Coursaris, C., & Léger, P. M. (2020, July). Visualizing a User's Cognitive and Emotional Journeys: A Fintech Case. In International Conference on Human-Computer Interaction (pp. 549-566). Springer, Cham.

Wang, Q., Yang, S., Liu, M., Cao, Z., & Ma, Q. (2014). An eye-tracking study of website complexity from cognitive load perspective. Decision support systems, 62, 1-10.

Wang, Y., Yu, S., Ma, N., Wang, J., Hu, Z., Liu, Z., & He, J. (2020). Prediction of product design decision Making: An investigation of eye movements and EEG features. Advanced Engineering Informatics, 45, 101095.

Wood, C., Torkkola, K., & Kundalkar, S. (2004). Using driver's speech to detect cognitive workload. In *9th Conference Speech and Computer*.

Wu, C. H., Liu, C. H., & Huang, Y. M. (2022). The exploration of continuous learning intention in STEAM education through attitude, motivation, and cognitive load. International Journal of STEM Education, 9(1), 1-22.

Xu, N., Guo, G., Lai, H., & Chen, H. (2018). Usability Study of Two In-Vehicle Information Systems Using Finger Tracking and Facial Expression Recognition Technology. International Journal of Human-Computer Interaction, 34(11), 1032–1044

Yin, B., Chen, F., Ruiz, N., & Ambikairajah, E. (2008). Speech-based cognitive load monitoring system. In 2008 IEEE International Conference on Acoustics, Speech and Signal Processing (pp. 2041–2044). IEEE. https://doi.org/10.1109/ICASSP.2008.4518041

Zagermann, J., Pfeil, U., & Reiterer, H. (2016, October). Measuring cognitive load using eye tracking technology in visual computing. In Proceedings of the sixth workshop on beyond time and errors on novel evaluation methods for visualization (pp. 78-85).

Zardari, B.A., Hussain, Z., Arain, A.A.  QUEST e-learning portal: applying heuristic evaluation, usability testing and eye tracking.  20, 531–543 (2021). https://doi.org/10.1007/s10209-020-00774-z

Zheng, R. Z. (Ed.). (2017). Cognitive load measurement and application: a theoretical framework for meaningful research and practice. Routledge.

Zhou, T., Cha, J. S., Gonzalez, G., Wachs, J. P., Sundaram, C. P., & Yu, D. (2020). Multimodal physiological signals for workload prediction in robot-assisted surgery. ACM Transactions on Human-Robot Interaction (THRI), 9(2), 1-26.